# 3D WC-MPS coupled with geometrically nonlinear shell for hydro-elastic free-surface flows

**Rubens Augusto Amaro Junior**
Department of Construction Engineering, Polytechnic School of University of São Paulo
Av. Prof. Almeida Prado, trav. 2, 83 - Cidade Universitária, 05508-070, São Paulo, SP, Brazil
e-mail: rubens.amaro@usp.br
**Alfredo Gay Neto**
Department of Structural and Geotechnical Engineering, Polytechnic School of University of São Paulo
Av. Prof. Almeida Prado, trav. 2, 83 - Cidade Universitária, 05508-070, São Paulo, SP, Brazil
e-mail: alfredo.gay@usp.br
**Liang-Yee Cheng**
Department of Construction Engineering, Polytechnic School of University of São Paulo
Av. Prof. Almeida Prado, trav. 2, 83 - Cidade Universitária, 05508-070, São Paulo, SP, Brazil
e-mail: cheng.yee@usp.br

**Abstract:** A 3D fluid-structure interaction solver based on an improved weakly-compressible moving particle simulation (WC-MPS) method and a geometrically nonlinear shell structural model is developed and applied to hydro-elastic free-surface flows. The fluid-structure coupling is performed by a polygon wall boundary model that can handle particles and finite elements of distinct sizes. In WC-MPS, a tuning-free diffusive term is introduced to the continuity equation to mitigate non-physical pressure oscillations. Discrete divergence operators are derived and applied to the polygon wall boundary, of which the numerical stability is enhanced by a repulsive Lennard-Jones force. Additionally, an efficient technique to deal with the interaction between fluid particles placed at opposite sides of zero-thickness walls is proposed. The geometrically nonlinear shell is modeled by an unstructured mesh of six-node triangular elements. Finite rotations are considered with Rodrigues parameters and a hyperelastic constitutive model is adopted. Benchmark examples involving free-surface flows and thin-walled structures demonstrate that the proposed model is robust, numerically stable and offers more efficient computation by allowing mesh size larger than that of fluid particles.

**Key words:** Hydroelasticity, Free-surface flow, Nonlinear shell formulation, Particle-based method, Finite element method (FEM)

## 1 Introduction

In most engineering applications, the design of mechanical systems by considering fluid-structure interaction (FSI) phenomena is of great importance due to safety, reliability, efficiency, or durability issues. Among many phenomena involving fluid and structure motions in extreme conditions, slamming/green-water on ships, sloshing in liquid storage systems, tsunami/storm/flood impact on structures, are examples characterized by violent free-surface flows with impulsive hydrodynamic loads. With the remarkable advances in high-performance computing (HPC) systems, numerical simulations are promising approach to deal with such kind of problems.

The mesh-free particle-based (Lagrangian) methods, such as smoothed particle hydrodynamics (SPH) [1, 2] and moving particle semi-implicit/simulation (MPS) [3], are very effective to model violent FSI free-surface flows with large interfacial deformation and fragmentations [4, 5]. These methods have been widely developed in the context of hydroelasticity [6, 7] and many others as reviewed by Gotoh et al. [8], with some focused on parallel computing in graphics processing units (GPU) [9, 10], and applied to solve related problems such as sloshing [11, 12], slamming [13], dam breaking [14], tsunami [15, 16], structure failures [17, 18, 19], composite structures [20] and biomedical engineering [21].

Nevertheless, since uniform spatial resolution is usually adopted in the fluid domain for large-scale FSI problems involving complex or thin structures, huge computational efforts associated with high-resolution models are demanded. In that case, some authors adopted multi-resolution techniques [22, 23, 24, 25, 26, 27], in which high-resolution is used only near the structures or local critical areas. However, these formulations are generally complex, especially for three-dimensional (3D) problems, and demand a considerable coding implementation effort.

Meanwhile the particle-based methods can model highly nonlinear violent free-surface flows and multi-bodies with complex geometries, from the solid (structural) point of view, the finite element method (FEM) can handle a great variety of solid and structural models. Thus, recently, many researchers have taken benefit from meshless particle-based methods along with mesh-based methods to develop coupled FSI solvers.

In this direction, SPH was coupled to FEM solvers by De Vuyst [28] using a particle-particle contact approach, by Groenenboom and Cartwright [29] adopting sliding interface algorithms based on the penalty formulation, and by Yang et al. [30] with the Monaghan repulsive boundary condition [31, 32] applied to wall-particle interface. Li et al. [33] proposed a coupled

SPH-ALE (arbitrary Lagrangian–Eulerian) [34] and FE methods considering geometrically linear structures, and further allowing large structural deformation [35]. Next, to optimize the numerical efficiency and stability, a multiple-time-step algorithm proposed by Gravouil and Combescure [36] was incorporated by Nunez-Ramirez et al. [37], allowing different time steps in each sub-domain. Nevertheless, the techniques adopted in Li et al. [33, 35] and Nunez-Ramirez et al. [37] are not generic since they are strongly linked to the FE model, and considerable coding implementation effort within the solid solver is required. All aforementioned works adopted elements and particles of equal or almost the same size.

Long et al. [38] introduced a 2D particle-element contact algorithm based on master-slave scheme into the coupling of FEM-ISPH (incompressible SPH) [39, 40] and FEM-SPH. Further, they proposed a numerical technique where ghost particles are dynamically generated to compensate the truncate compact support of SPH particles near solid elements [41]. Similarly, in Fuchs et al. [42], the truncated support is completed by a set of virtual boundary particles, which slide along the fluid-structure interface following the motion of the fluid particles.

Concerning coupling algorithms, Fourey et al. [43] compared the conventional parallel staggered (CPS) procedure for parallel algorithm, in which both the SPH and FEM solvers update synchronously the solution to the next time step, and the conventional sequential staggered (CSS) for sequential algorithm, in which the solvers progress alternatively [44]. They concluded that CSS is more stable than CPS, although CPS shows a better performance in terms of CPU time saving. Later, Hermange et al. [45] extended the formalism developed in Fourey et al. [43] to a 3D model and applied it to hydroplaning phenomenon. Ogino et al. [46] developed a partitioned coupling of SPH-FE methods by adopting interface marker on fluid-structure boundaries and a dummy mesh for the fluid domain. For more detail description about the numerical modeling of the interaction between particles and finite elements, the readers are invited to peruse the work of Liu and Zhang [47]. Moreover, for the practical applications of SPH-FE models, Groenenboom et al [48, 49] might also be mentioned.

Regarding the efforts on the coupling of MPS and FE methods to deal with FSI problems, by authors' knowledge the earliest attempt was the work of Lee et al. [50]. They adopted the same resolution for both fluid particles and 2D geometrically nonlinear shell elements, which were weakly coupled by pseudo fluid particles placed at the FE nodes. Similarly, Mitsume et al. [51] coupled 2D explicit MPS (EMPS) [52] and FE methods by overlapping wall boundary particles and FE nodes for the data exchange in fluid-structure interfaces. Subsequently, Mitsume et al. [53] improved their model by using a polygon wall boundary model [54] to tackle 2D complex shaped fluid-structure boundaries.

Rao et al. [55] proposed a weakly coupled MPS-FEM to investigate the interaction between a solitary waves and elastic structures in 2D. Afterwards, Zhang and Wan [56] improved the 2D coupled model by including a particle group scheme [11], in which solid particles located within a transverse section are grouped. The hydrodynamic forces of the sections are applied on the corresponding structural beam nodes, of which the displacements are computed by FEM and feedbacked to the corresponding particle group. In Zhang and Wan [57] and Chen et al. [58] the model was extended to solve 3D problems. Further, Zhang et al. [59] compared two data interpolation techniques, namely shape function based interpolation technique (SFBI) and kernel function-based interpolation (KFBI), concluding that numerical stability and robustness are improved by SFBI, while a better accuracy is given by KFBI. A partitioned one-way coupling of EMPS-FEM that neglects the reaction (displacement) of the solid on fluid was reported in Zheng et al. [60] and Zheng and Shioya [61]. Recently, Zheng et al. [62] proposed a 3D coupled EMPS-FEM by using ghost cell boundary (GCB) model, in which integration points of cells are distributed to solid finite elements for data exchange in the interface.

As previously reported, most of the existing works adopt solid finite elements of size restricted by the particle scale. Since the required model complexity and forthcoming number of degrees of freedom may be drastically distinct in solids and fluids, independent resolutions in both domains are desirable. In addition to solid and rod, FEM can also be used to model shell elements, which are particularly useful to simulate thin-walled structures experiencing bending/membrane effects. Furthermore, the adoption of shell elements of arbitrary size has the potential to significantly save the computational cost since they allow the modeling of thin-walled structures with fewer elements. Despite the 2D implementation using geometrically nonlinear shell elements of size restricted by fluid particles dimension adopted by Lee et al. [50] to the best knowledge of the authors, there is almost no study reporting 3D coupled particle-mesh models, at least in the context of SPH and MPS, adopting geometrically nonlinear shell structures and able to handle finite elements and particles of distinct sizes.

Under these circumstances, our main objective is to propose a FSI solver, by coupling an improved particle-based method and a robust geometrically exact shell model to handle nonlinear transient applications in the presence of free-surface flow and structures subjected to geometric nonlinearity. Here, the in-house 3D code based on the proposed improved weakly-compressible MPS method (WC-MPS)[1], which is also an explicit solver [63], and the nonlinear FE solver Giraffe (Generic Interface Readily Accessible For Finite Elements) [64] are coupled adopting the explicitly represented polygon (ERP) wall boundary model in such a way that the resolution of the solid mesh is not constrained by the particle size.

[1] *The WC-MPS source code can be found at https://github.com/rubensamarojr/polymps*

Among the improvements proposed for WC-MPS, we have the calculation of particle number density using the continuity equation with introduction of a tuning-free diffusive term, inspired on the SPH formulation of [65], that mitigates the numerical noise on the pressure field, which are harmful for the dynamic coupling of fluid-structural solver [66], and significantly improves the smoothness and the accuracy of pressure field compared to the original WC-MPS or EMPS methods.

As another original contribution, which also significantly improves the numerical stability, is the introduction of a repulsive force based on Lennard-Jones potential [31] in the ERP formulation. In addition, new discrete divergence operators are herein derived and applied for the formalism of ERP. Finally, a simple but effective technique is proposed to avoid the incorrect interaction between fluid particles placed at opposite sides of zero-thickness polygon walls.

Numerical examples of interaction between free-surface flows and thin-walled structures demonstrate that the proposed model is robust and numerically stable. Moreover, the effect of the relation between mesh size and particle distance were also investigated, showing that the ability of dealing with mesh size larger than fluid particle distance offers more efficient computation.

# 2 Numerical methods

## 2.1 Fluid domain

The governing equations for a weakly-compressible, barotropic fluid can be written in a moving Lagrangian frame as:

$$\begin{cases} \dfrac{D\rho_f}{Dt} = -\rho_f \nabla \cdot \mathbf{u} \\ \dfrac{D\mathbf{u}}{Dt} = -\dfrac{\nabla P}{\rho_f} + \nu_f \nabla^2 \mathbf{u} + \mathbf{f}_e \\ \dfrac{D\mathbf{r}}{Dt} = \mathbf{u} \\ P = F(\rho_f) \end{cases}, \tag{1}$$

where $\rho_f$ is the fluid density, $\mathbf{u}$ denotes the velocity vector, $P$ represents the pressure, $\nu_f$ stands for the fluid kinematic viscosity, $\mathbf{f}_e$ is the vector of external body force per unit mass, $\mathbf{r}$ means the position vector and $F$ is the state equation linking the pressure and density.

In MPS, the differential operators are replaced by discrete operators $\langle \quad \rangle_i$ on a $i$-th target particle, derived from a weight function $\omega_{ij}$ that accounts for the influence of its neighbor particle $j$ inside the region limited by the effective radius $r_e = kl_0$, with $k \in [2.0, 4.0]$ [3] and $l_0$ the initial particle distance. Here we adopted the rational weight function given by [3]:

$$\omega_{ij} = \begin{cases} \left(\frac{r_{\mathbf{e}}}{\|\mathbf{r}_{ij}\|} - 1\right) & \|\mathbf{r}_{ij}\| \le r_e \\ 0 & \text{otherwise} \end{cases}, \tag{2}$$

where $\mathbf{r}_{ij} = \mathbf{r}_j - \mathbf{r}_i$ and $\|\mathbf{r}_{ij}\|$ stands the distance between the particles $i$ and $j$.

### 2.1.1 Particle number density

The fluid density $\rho_i$ can be related to the particle number density $n_i$ as [67]:

$$\frac{n_i}{n^0} = \frac{\rho_i}{\rho^0}, \tag{3}$$

where $n^0$ stands for the constant particle number density computed considering a fully compact support with an initial cubic arrangement of particles and $\rho^0$ is the reference density. The particle number density $n_i$ was originally defined as:

$$n_i = \sum_{j \in \mathbb{P}_i} \omega_{ij}, \tag{4}$$

where $\mathbb{P}_i$ is the set of neighboring particles of the particle $i$.

### 2.1.2 Explicit algorithm

The numerical integration of the momentum equation (1) is evaluated with an explicit predictor-corrector scheme. At first, predictions of the velocity and position for a $i$-th fluid particle are obtained by using viscosity and external forces terms:

$$\mathbf{u}_i^* = \mathbf{u}_i^t + \left(\nu_f \langle \nabla^2 \mathbf{u} \rangle_i^t + \mathbf{f}_e^t\right) \Delta t, \tag{5}$$

$$\mathbf{r}_i^* = \mathbf{r}_i^t + \mathbf{u}_i^* \, \Delta t, \tag{6}$$

where the superscript * refers to an intermediate value at the prediction step. The Laplacian of velocity is approximated by [3]:

$$\langle \nabla^2 \mathbf{u} \rangle_i^t = \frac{2d}{\lambda^0 n^0} \sum_{j \in \mathbb{P}_i} \left(\mathbf{u}_j^t - \mathbf{u}_i^t\right) \omega_{ij}^t, \tag{7}$$

where $d = 1, 2$ or $3$ is the number of spatial dimensions, and $\lambda^0$ refers to a correction parameter:

$$\lambda^0 = \frac{\sum_{j\in\mathbb{P}_i}\left\|\mathbf{r}_{ij}^0\right\|^2\omega_{ij}^0}{\sum_{j\in\mathbb{P}_i}\omega_{ij}^0}. \tag{8}$$

After that, to avoid the clustering of fluid particles, a pair-wise collision (PC) model [68] is applied, i.e., the collision contribution $\Delta\mathbf{u}_i^*$ reads:

$$\Delta\mathbf{u}_i^* = \begin{cases} \sum_{j\in\mathbb{P}_i}\frac{(1+\alpha_r)}{2}\frac{\mathbf{r}_{ij}^*\cdot\mathbf{u}_{ij}^*}{\left\|\mathbf{r}_{ij}^*\right\|}\frac{\mathbf{r}_{ij}^*}{\left\|\mathbf{r}_{ij}^*\right\|} & \left\|\mathbf{r}_{ij}^*\right\| \le \alpha_d l_0 \quad \text{and} \quad \mathbf{r}_{ij}^*\cdot\mathbf{u}_{ij}^* < 0 \\ 0 & \text{otherwise} \end{cases}. \tag{9}$$

Lee et al. [68], [68] investigated different combinations of the collision coefficients and found the optimal ranges of coefficient of restitution $\alpha_r \in ]0.,0.2]$ and collision distance $\alpha_d \in [0.8, 1[$ that improve the spatial stability. Based on previous works using particle [69, 70] or polygon [71, 72] wall models, and our experience, the adoption of $\alpha_r \approx 0.2$ and collision distance $\alpha_d \approx 0.9$ ensures stable numerical simulations for most of the problems. In this way, we adopted the coefficients $\alpha_r = 0.2$ and $\alpha_d = 0.85$ for all simulations.

In order to couple the continuity and momentum equations, Eq. (1), the Tait's equation of state (EOS) is explicitly calculated providing the pressure of a $i$-th fluid particle [63]:

$$P_i^{t+\Delta t} = \frac{\rho_f c_0^2}{\gamma}\left[\left(\frac{n_i^*}{n^0}\right)^\gamma - 1\right], \tag{10}$$

where $c_0 = \sqrt{\partial P/\partial\rho}$ stands for the speed of sound in the reference density, the polytrophic index $\gamma = 7$ is a typical value adopted for fluid phase and $n_i^*$ is the particle number density calculated after the prediction process, here computed as $n_i^* = n_i^t + \Delta n_i^*$ with $n_i^t$ estimated at the previous step $t$ and $\Delta n_i^*$ given by Eq. (19). An artificial $c_0$ (smaller than physical one) is usually adopted to prevent numerical instabilities and avoid extremely small $\Delta t$, see Monaghan [31].

Afterwards, the velocity of the fluid particles ($\mathbf{u}_i^{t+\Delta t}$) is updated and the positions $\mathbf{r}_i'$ are corrected, both through a simple first-order Euler integration:

$$\mathbf{u}_i^{t+\Delta t} = \mathbf{u}_i^* + \Delta\mathbf{u}_i^* - \frac{\Delta t}{\rho_f}\langle\nabla P\rangle_i^*, \tag{11}$$

$$\mathbf{r}_i' = \mathbf{r}_i^* + \left(\mathbf{u}_i^{t+\Delta t} - \mathbf{u}_i^*\right)\Delta t. \tag{12}$$

Here, we adopted an antisymmetric momentum formulation for the pressure gradient [73]:

$$\langle\nabla P\rangle_i^* = \frac{d}{n^0}\sum_{j\in\mathbb{P}_i}\left(\frac{n_i^*}{n_j^*}P_j^{t+\Delta t} + \frac{n_j^*}{n_i^*}P_i^{t+\Delta t}\right)\frac{\mathbf{r}_{ij}^*}{\left\|\mathbf{r}_{ij}^*\right\|^2}\omega_{ij}^*. \tag{13}$$

Finally, to reduce the discrepancy between the particle number densities $n_i^*$ and $n^0$, particle shifting (PS) governed by Fick's law of diffusion [74] is adopted:

$$\Delta \mathbf{r}_i' = -A_F (l_0)^2 C_r Ma \begin{cases} \langle \nabla C \rangle_i' & i \in \mathbb{I} \\ 0 & i \in \mathbb{F} \end{cases}, \tag{14}$$

where $i \in \mathbb{I}$ represents the inner fluid particles, $i \in \mathbb{F}$ is the free surface particles, $A_F$ is a constant and $Ma$ denotes the Mach number.

The magnitude of the $A_F$ must be in a range of values such that the PS improves the numerical stability, but does not introduce significant errors. In this sense, the range $A_F \in [1, 6]$ was tested and found to work satisfactorily in Skillen et al. [75]. Moreover, the value of $A_F = 2$ has been found to provide a good compromise for SPH [75] and MPS [76] simulations. Heuristically, we observed that $A_F = 1$ increases the numerical dissipation, resulting in a reduced flow velocity in MPS simulations. In this way, in order to avoid significant motion of fluid particles, we adopted $A_F = 1$ for the hydrostatic test. On the other hand, since $A_F = 2$ has a less influence on the numerical dissipation but ensures more ordered particle distribution, this value was adopted for all dynamic simulations.

The gradient of the concentration $C_i$ (volume fraction) in Eq. (14) can be calculated as [73]:

$$\langle \nabla C \rangle_i = \frac{d}{n^0} \sum_{j \in \mathbb{P}_i} \frac{C_i + C_j}{\|\mathbf{r}_{ij}\|^2} \mathbf{r}_{ij} \omega_{ij} \ . \tag{15}$$

with

$$C_i = \frac{\sum_{j \in \mathbb{P}_i} \omega_{ij}}{n^0}. \tag{16}$$

At the end of each time step, $\Delta \mathbf{r}_i'$ is imposed to inner fluid particles $i \in \mathbb{I}$, i.e., no free-surface, and the final position is adjusted as:

$$\mathbf{r}_i^{t+\Delta t} = \mathbf{r}_i' + \Delta \mathbf{r}_i' \ . \tag{17}$$

For all simulations, a fixed time step based on the maximum flow velocity $|u|_{\max}$, was initially assigned following the CFL condition [77]:

$$\Delta t \leq \frac{C_r \, l_0}{|u|_{\max}}, \tag{18}$$

where $C_r$ denotes the Courant number.

#### 2.1.3 Improved WC-MPS with free-tuning diffusive term (CD-WC-MPS)

For fully explicit methods, and with less influence on implicit methods, Eq. (4) leads to noisy density estimation because it is more sensitive to small changes of the local particles distribution [78]. As a result, non-physical spurious oscillations of pressure, which is computed directly from $n_i$, is obtained. To overcome this issue, $n_i$ is obtained herein by substituting Eq. (3) into the discrete form of the continuity equation (1) including a diffusive term $D_i$, similar to that originally proposed by Molteni and Colagrossi [66] on the so-called $\delta$-SPH method, and recently adapted for WC-MPS in Jandaghian and Shakibaeinia [73]:

$$\frac{1}{n_i}\frac{\Delta n_i}{\Delta t} = -\langle \nabla \cdot \mathbf{u} \rangle_i + D_i \, , \tag{19}$$

where $\Delta t$ denotes the time-step. Notwithstanding, both formulations [66 73] incorporate a tuning parameter which requires calibration. Eq. (19) is used only to update the $n_i^*$ in EOS (see Eq. (10)). Equivalent diffusive terms involving the Laplacian of density or Laplacian of pressure can be found in the literature [79].

Inspired by the work of Fernández-Gutiérrez and Zohdi [65] about SPH, we adopted a free-tuning diffusive term $D_i$ for the WC-MPS, hereinafter referred to as CD-WC-MPS, as follows:

$$D_i = \frac{\Delta t}{\rho_f}\langle \nabla^2 \bar{P} \rangle_i = \frac{\Delta t}{\rho_f}\frac{2d}{\lambda^0 n^0}\sum_{j \in \mathbb{P}_i}\left[P_j - P_i - \left(-\rho_f \mathbf{g} \cdot \mathbf{r}_{ij}\right)\right]\omega_{ij} \, . \tag{20}$$

As explained in Fernández-Gutiérrez and Zohdi [65], the diffusive term in Eq. (20) reduces some of the spurious numerical high-frequency oscillations in the pressure field, and therefore suppresses the induced sound wave and smoothes the calculated particle number density field. Moreover, such diffusive term presents a good consistency related to the continuity equation (mass conserving equation), and thermodynamic conservation (energy balance), although global mass conservation at a discrete level is not ensured, as verified in Cercos-Pita et al. [79].

Fortakas et al. [80] demonstrated that for gravity-dominated flows the computed pressure close to the boundaries can be improved by removing the hydrostatic contribution from the total density in the diffusive term. Simillarly, a term including the gravity vector is added in Eq. (20) to remove the expected pressure difference from the hydrostatic forces [65].

With respect to the divergence of the velocity in Eq. (19), it is approximated by [73]:

$$\langle \nabla \cdot \mathbf{u} \rangle_i = \frac{d}{n^0}\sum_{j \in \mathbb{P}_i}\left(\frac{n_j}{n_i}\right)\frac{\left(\mathbf{u}_j - \mathbf{u}_i\right)\cdot\left(\mathbf{r}_j - \mathbf{r}_i\right)}{\left\|\mathbf{r}_j - \mathbf{r}_i\right\|^2}\omega_{ij} \, . \tag{21}$$

## 2.2 Fluid domain boundaries

### 2.2.1 Free surface

In the particle-based methods, the kinematic boundary condition of a free surface is directly satisfied by the motion of the free-surface particles, while the dynamic boundary condition is imposed by applying the Dirichlet condition $P = 0$ at free-surface particles.

The present study adopts the neighborhood particles centroid deviation (NCPD) technique [81] to identify free surface particles $\mathbb{F}$. In the first step, the free surface particle detection based on particle number density criterion is carried out:

$$\begin{cases} \sum_{j \in \mathbb{P}_i} \omega_{ij} < \beta_{FS} \cdot n^0 & \rightarrow \quad i \in \mathbb{F} \\ \text{otherwise} & \rightarrow \quad i \in \mathbb{I} \end{cases} . \tag{22}$$

In the second step of NPCD, criterion of centroid deviation is applied only on free-surface particles to eliminate the misdetection of inner particles as the free-surface ones that occurred in the first step:

$$\begin{cases} N_i < 4 \text{ or } \sigma_i > \varrho_{FS} \cdot l_0 & \rightarrow \quad i \in \mathbb{F} \\ \text{otherwise} & \rightarrow \quad i \in \mathbb{I} \end{cases} , \tag{23}$$

with the constants $\beta_{FS} \in [0.8, 1[$ [3] and $\varrho_{FS} \in [0.2, \infty[$ [81]. The value $N_i$ represents the number of neighboring particles $j \in \mathbb{P}_i$, and the deviation $\sigma_i$ is calculated as:

$$\sigma_i = \frac{\left\| \sum_{j \in \mathbb{P}_i} (\omega_{ij} \mathbf{r}_{ij}) \right\|}{\sum_{j \in \mathbb{P}_i} \omega_{ij}} = \frac{\sqrt{\left[\sum_{j \in \mathbb{P}_i} \omega_{ij} (x_j - x_i)\right]^2 + \left[\sum_{j \in \mathbb{P}_i} \omega_{ij} (y_j - y_i)\right]^2 + \left[\sum_{j \in \mathbb{P}_i} \omega_{ij} (z_j - z_i)\right]^2}}{\sum_{j \in \mathbb{P}_i} \omega_{ij}} . \tag{24}$$

### 2.2.2 Explicitly represented polygon (ERP) wall boundary model

Instead of the conventional MPS formulation that represents the wall boundaries by discrete layers of wall and dummy (ghost) particles, see Fig. 2-1(a), here we adopted the explicitly represented polygon (ERP) wall boundary model [82]. The ERP represents solid boundaries as triangular polygons, see Fig. 2-1(b), which are explicitly represented without using the signed distance function (SDF). Moreover, it is assumed that each fluid particle is affected by only the closest triangular polygon. Here, an axis-aligned bounding box (AABB) hierarchy, further explained in section 2.2.3.4, is adopted to find the closest triangule from a particle.

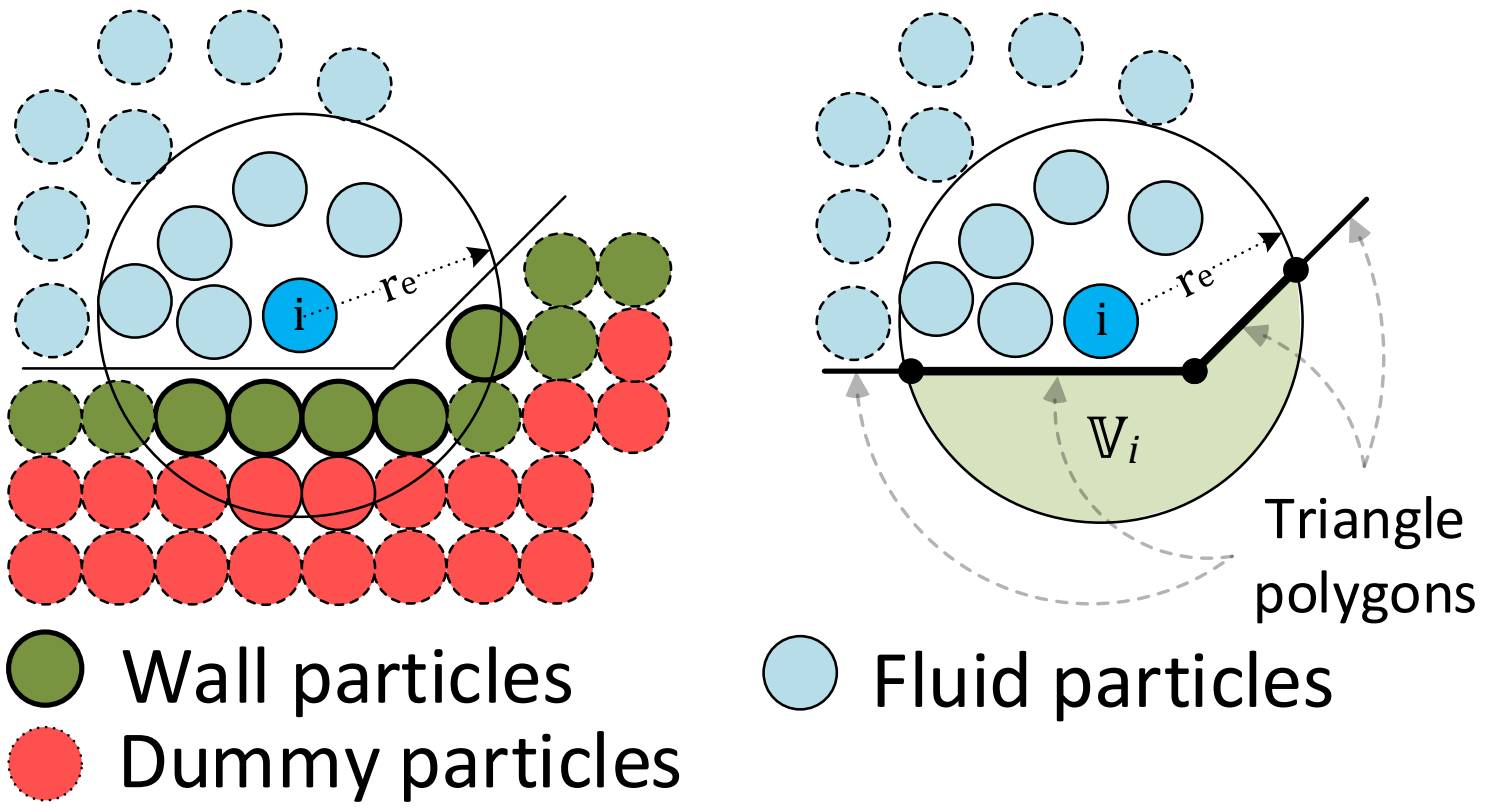

(a) (b)

Fig. 2-1. 2D schematic view of fluid particles interacting with the (a) conventional wall particle model and (b) polygon wall boundary model.

Since the compact support of particles near the polygon walls is not fully filled, i.e. truncated support as showed in Fig. 2-1(b), their numerical operators are divided into the contribution due to neighbor fluid particles $\langle\ \ \rangle_{j\in\mathbb{P}_{i\to i}}$, in the same way of the standard WC-MPS, and the virtual neighboring particles $\mathbb{V}_i$, represented by the closest polygon wall $\langle\ \ \rangle_{\mathrm{wall}\to i}$, as illustrated in Fig. 2-2. To calculate the numerical operators $\langle\ \ \rangle_{\mathrm{wall}\to i}$, first, the position of the mirror particle $i'$ corresponding to particle $i$ is computed (Eq. (30)). After that, the numerical operators of particle $i'$ are calculated considering all the particles inside its neighbor region, i.e., the original particle $i$ and its neighbors, see Fig. 2-2(b). Finally, these operators are multiplied by a transformation matrix ($\mathbf{R}_i^{\mathrm{ref}}$ or the identity matrix $\mathbf{I}$) and added to the numerical operators $\langle\ \ \rangle_{j\in\mathbb{P}_{i\to i}}$, see Fig. 2-2(c).

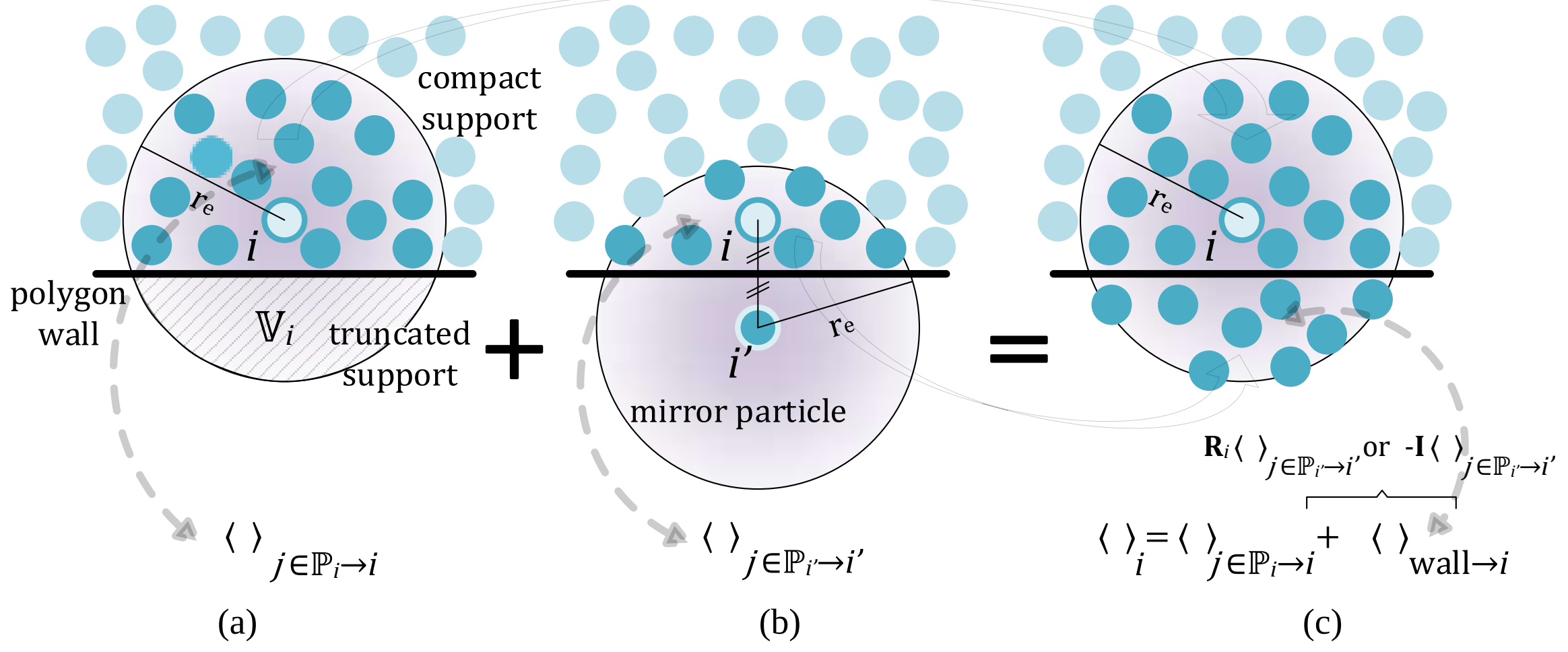

Fig. 2-2. ERP model. Contribution of fluid particles and polygon walls to the calculation of numerical operators.

The pressure gradient and the Laplacian of the velocity terms of fluid particles close to polygon wall boundary are calculated as:

$$\langle\nabla P\rangle_i = \langle\nabla P\rangle_{j\in\mathbb{P}_{i\to i}} + \langle\nabla P\rangle_{\mathrm{wall}\to i}\,, \tag{25}$$

$$\langle\nabla^2\mathbf{u}\rangle_i = \langle\nabla^2\mathbf{u}\rangle_{j\in\mathbb{P}_{i\to i}} + \langle\nabla^2\mathbf{u}\rangle_{\mathrm{wall}\to i}\,. \tag{26}$$

We adapted the pressure gradient operator proposed by Mitsume et al. [82], following Jandaghian and Shakibaeinia [73], and the operator $\langle\ \ \rangle_{\mathrm{wall}\to i}$ for pressure gradient reads:

$$\langle\nabla P\rangle_{\mathrm{wall}\to i} = \mathbf{R}_i^{\mathrm{ref}}\frac{d}{n^0}\sum_{j\in\mathbb{P}_{i'}}\left(\frac{n_i}{n_j}P_j + \frac{n_j}{n_i}P_i\right)\frac{\mathbf{r}_{i'j}}{\left\|\mathbf{r}_{i'j}\right\|^2}\omega_{i'j}\,, \tag{27}$$

where $\mathbf{r}_{i'j} = \mathbf{r}_j - \mathbf{r}_{i'}$. The operators $\langle\ \ \rangle_{\mathrm{wall}\to i}$ for the Laplacian of the velocity for the free-slip (Eq. (28)) and no-slip boundary (Eq. (29)) condition are computed via a weighted sum over neighboring fluid particles $\mathbb{P}_{i'}$ of the mirror particle $i'$ [82]:

$$\langle \nabla^2 \mathbf{u} \rangle_{\text{wall}\to i} = \mathbf{R}_i^{\text{ref}} \frac{2d}{\lambda^0 n^0} \sum_{j\in\mathbb{P}_{i'}} \left(\mathbf{u}_j - \mathbf{u}_{i'}\right)\omega_{i'j} \quad , \quad \mathbf{u}_{i'} = \mathbf{R}_i^{ref} \mathbf{u}_i \, , \tag{28}$$

$$\langle \nabla^2 \mathbf{u} \rangle_{\text{wall}\to i} = -\mathbf{I} \frac{2d}{\lambda^0 n^0} \sum_{j\in\mathbb{P}_{i'}} \left(\mathbf{u}_j - \mathbf{u}_{i'}\right)\omega_{i'j} \quad , \quad \mathbf{u}_{i'} = -\mathbf{u}_i + 2\left[\mathbf{u}_i^{\text{wall}} - \left(\mathbf{n}_i^{\text{wall}} \cdot \mathbf{u}_i^{\text{wall}}\right)\mathbf{n}_i^{\text{wall}}\right] \tag{29}$$

where $\mathbf{u}_i^{\text{wall}}$ is the velocity of the wall at the midpoint of the line segment connecting the particles $i - i'$, and $\mathbf{n}_i^{\text{wall}}$ is the unit normal vector of the wall pointing to the particle $i$.

The position of the mirror particle $i'$ corresponding to particle $i$ is obtained as:

$$\mathbf{r}_{i'} = \mathbf{r}_i + 2\left(\mathbf{r}_i^{\text{wall}} - \mathbf{r}_i\right) , \tag{30}$$

where $\mathbf{r}_i^{\text{wall}}$ is the closest point on the polygon to particle $i$. Let $\mathbf{r}_{i,\mathbb{T}}^{\text{close}}$ be the closest points on each triangle polygon $\mathbb{T}$ from the $i$-th particle, the closest point $\mathbf{r}_i^{\text{wall}}$ is defined by:

$$\mathbf{r}_i^{\text{wall}} = \underset{\mathbf{r}_{i,\mathbb{T}}^{\text{close}}}{\operatorname{argmin}} \left\| \mathbf{r}_i - \mathbf{r}_{i,\mathbb{T}}^{\text{close}} \right\| \quad , \quad \mathbb{T} \in [1, 2, \dots, N_{\mathbb{T}}] \, , \tag{31}$$

where $N_{\mathbb{T}}$ is the number of triangle polygons.

The transformation matrix $\mathbf{R}_i^{\text{ref}}$ for reflection across the plane is expressed as [83]:

$$\mathbf{R}_i^{\text{ref}} = \mathbf{I} - 2\mathbf{n}_i^{\text{wall}} \otimes \mathbf{n}_i^{\text{wall}} , \tag{32}$$

where $\mathbf{I}$ denotes the identity matrix.

The contribution of the closest polygon wall on the gradient of the concentration (Eq. (15)) can be approximated by:

$$\langle \nabla C \rangle_{\text{wall}\to i} = \mathbf{R}_i^{\text{ref}} \frac{d}{n^0} \sum_{j\in\mathbb{P}_{i'}} \frac{C_i + C_j}{\left\| \mathbf{r}_{i'j} \right\|^2} \mathbf{r}_{i'j} \omega_{i'j} \, . \tag{33}$$

Since $\mathbf{r}_{ij'} = \mathbf{R}_i^{\text{ref}} \mathbf{r}_{i'j}$, the deviation $\sigma_i$ (see Eq. (24)) of fluid particles close to polygon walls is rewritten as:

$$\sigma_i = \frac{\left\| \sum_{j\in\mathbb{P}_i} \left(\omega_{ij} \mathbf{r}_{ij}\right) + \mathbf{R}_i^{\text{ref}} \sum_{j\in\mathbb{P}_{i'}} \left(\omega_{i'j} \mathbf{r}_{i'j}\right) \right\|}{\sum_{j\in\mathbb{P}_i} \omega_{ij} + \sum_{j\in\mathbb{P}_{i'}} \omega_{i'j}} \, . \tag{34}$$

The particle number density variation $\Delta n_i$ also is partitioned into the contribution due to the fluid particles $\Delta n_{\mathbb{P}_i \to i}$, see Eq. (19), and the polygon walls $\Delta n_{\text{wall}\to i}$, i.e., $\Delta n_i = \Delta n_{\mathbb{P}_i \to i} + \Delta n_{\text{wall}\to i}$.

Instead of the formulation of Eq. (19), the density variation due to the polygon walls $\Delta n_{\text{wall}\to i}$ is calculated without applying $D_i$:

$$\frac{1}{n_i} \frac{\Delta n_{\text{wall}\to i}}{\Delta t} = -\langle \nabla \cdot \mathbf{u} \rangle_{\text{wall}\to i} \, . \tag{35}$$

This is because the diffusive term $D_i$ in Eq. (19) is required only for the improvement of the fluid particles distribution. Moreover, since the divergence of the velocity over the polygon wall boundary is required in Eq. (35), discrete divergence operators for ERP were derived in present work for both free-slip and no-slip boundary conditions.

The sum $\sum \omega_{ij}$ and the number of neighbors $N_i$ used in the free-surface detection, see Section 2.2.1, are computed considering the contribution due the fluid particles ($j \in \mathbb{P}_i \to i$) and the polygon walls (wall $\to i$). Under the assumption that the wall near the fluid particle is flat, the dummy particles $j'$ are initially arranged in a uniform particle distribution below the flat wall, and the wall contributions are evaluated as:

$$\left(\sum \omega_{ij}\right)_{\mathrm{wall}\to i} = f_1(\|\mathbf{r}_{iw}\|) \approx f_1(d_{iw}) \tag{36}$$

$$(N_i)_{\mathrm{wall}\to i} = f_2(\|\mathbf{r}_{iw}\|) \approx f_2(d_{iw})\,. \tag{37}$$

where $f_1$ and $f_2$ are determined by a linear interpolation of precomputed values at a given discrete distance $d_{iw}$, the normal distance between the $i$-th particle and the nearest polygon wall. It should be emphasized that $f_1$ and $f_2$ are computed at a few points within the effective radius $r_e$ at the beginning of the simulation and are stored in a lookup table, then saving the processing time.

### 2.2.3 Stability improvement of the wall repulsive force

To prevent penetrations of the free-surface particles $i \in \mathbb{F}$ into polygon walls or inner fluid particles $i \in \mathbb{I}$ at knuckled edges of polygon in case of curved surfaces (e.g., corners), a repulsive force $\mathbf{f}_i^{\mathrm{rep}}$, perpendicular to the boundary, is added to Eq. (27). Instead of the repulsive forces proposed by Mitsume et al. [82], with a specific coefficient, which should be tuned for each simulation, see e.g. [82, 84], or proposed by Harada [85], which is proportional to $1/\Delta t^2$, i.e., noticeably sensitive to any change on $\Delta t$, we introduced a repulsive force based on Lennard-Jones potential [31] in the ERP formulation:

$$\mathbf{f}_i^{\mathrm{rep}} = \begin{cases} -\dfrac{D_{\mathrm{rep}}}{\|\mathbf{r}_{iw}\|}\left[\left(\dfrac{0.5l_0}{\|\mathbf{r}_{iw}\|}\right)^{n1} - \left(\dfrac{0.5l_0}{\|\mathbf{r}_{iw}\|}\right)^{n2}\right]\mathbf{n}_i^{\mathrm{wall}} & \|\mathbf{r}_{iw}\| \le 0.5l_0 \\ 0 & \text{otherwise} \end{cases}, \tag{38}$$

where $(n_1, n_2)$ are integer coefficients usually taken as (12,4) or (4,2), $D_{\mathrm{rep}} = \rho C_{\mathrm{rep}}|V_{\mathrm{MAX}}|^2$ with $V_{MAX}$ the maximum velocity in the domain, the repulsive coefficient $C_{\mathrm{rep}} \in [1, 10]$ and $\mathbf{r}_{iw} = \mathbf{r}_i^{\mathrm{wall}} - \mathbf{r}_i$. In present work, the values $(n_1, n_2) = (4,2)$ and $C_{\mathrm{rep}} = 1$ are used for all simulations. Since $D_{\mathrm{rep}}$ is proportional to the instantaneous flow field, the Eq. (38) provides a

dynamic adjustment of the repulsive force, i.e., a little or no tuning of the coefficient $C_{\mathrm{rep}}$ is required, thereby significantly improving the numerical stability for a wide range of simulations.

### 2.2.4 Derivation of a new velocity divergence operator for free-slip boundary

Here, we extended the formulation proposed by Jandaghian and Shakibaeinia [73], see Eq. (21), to consider the polygonal walls. The divergence of the velocity associated with the virtual neighboring particles $\mathbb{V}_i$ (see Fig. 2-2), considering the free-slip boundary condition, is obtained by:

$$\langle \nabla \cdot \mathbf{u} \rangle_{\mathrm{wall}\to i} = \frac{d}{n^0} \sum_{j' \in \mathbb{V}_i} \left(\frac{n_j}{n_i}\right) \frac{\left(\mathbf{R}_j^{\mathrm{ref}} \mathbf{u}_j - \mathbf{u}_i\right) \cdot \mathbf{r}_{ij'}}{\left\|\mathbf{r}_{ij'}\right\|^2} \omega_{ij'} . \tag{39}$$

Assuming that the wall polygon nearest to the particle $i$ and its neighboring particle $j$ have the same unit normal vector $\mathbf{n}_j^{\mathrm{wall}} \approx \mathbf{n}_i^{\mathrm{wall}}$, one gets $\mathbf{R}_j^{\mathrm{ref}} \approx \mathbf{R}_i^{\mathrm{ref}}$, and Eq. (39) can be rewritten as:

$$\langle \nabla \cdot \mathbf{u} \rangle_{\mathrm{wall}\to i} = \frac{d}{n^0} \sum_{j \in \mathbb{P}_{i'}} \left(\frac{n_j}{n_i}\right) \frac{\left(\mathbf{R}_i^{\mathrm{ref}} \mathbf{u}_j - \mathbf{u}_i\right) \cdot \left(\mathbf{R}_i^{\mathrm{ref}} \mathbf{r}_{i'j}\right)}{\left\|\mathbf{r}_{i'j}\right\|^2} \omega_{i'j} . \tag{40}$$

Considering the orthogonal properties $\mathbf{R}_i^{\mathrm{ref}} = \left(\mathbf{R}_i^{\mathrm{ref}}\right)^T = \left(\mathbf{R}_i^{\mathrm{ref}}\right)^{-1}$ and $\mathbf{R}_i^{\mathrm{ref}} \mathbf{R}_i^{\mathrm{ref}} = \mathbf{I}$ of the transformation matrix, Eq. (40) yields:

$$\langle \nabla \cdot \mathbf{u} \rangle_{\mathrm{wall}\to i} = \frac{d}{n^0} \sum_{j \in \mathbb{P}_{i'}} \left(\frac{n_j}{n_i}\right) \frac{\mathbf{R}_i^{\mathrm{ref}}\left(\mathbf{u}_j - \mathbf{u}_{i'}\right) \cdot \left(\mathbf{R}_i^{\mathrm{ref}} \mathbf{r}_{i'j}\right)}{\left\|\mathbf{r}_{i'j}\right\|^2} \omega_{i'j} , \tag{41}$$

with $\mathbf{u}_{i'}$ provided in Eq. (28). Eq. (41) can be slightly simplified as:

$$\langle \nabla \cdot \mathbf{u} \rangle_{\mathrm{wall}\to i} = \frac{d}{n^0} \sum_{j \in \mathbb{P}_{i'}} \left(\frac{n_j}{n_i}\right) \frac{\left(\mathbf{u}_j - \mathbf{u}_{i'}\right) \cdot \mathbf{r}_{i'j}}{\left\|\mathbf{r}_{i'j}\right\|^2} \omega_{i'j} . \tag{42}$$

### 2.2.5 Derivation of a new velocity divergence operator for no-slip boundary

When the no-slip boundary condition is imposed on a wall, the velocity of the neighbor mirror particle $j'$ is defined by:

$$\mathbf{u}_{j'} = -\mathbf{u}_j + 2\left[\mathbf{u}_j^{\mathrm{wall}} - \left(\mathbf{n}_j^{\mathrm{wall}} \cdot \mathbf{u}_j^{\mathrm{wall}}\right)\mathbf{n}_j^{\mathrm{wall}}\right], \tag{43}$$

where $\mathbf{u}_j^{\mathrm{wall}}$ is the velocity of the wall at the midpoint of the line segment connecting the particles j–j' . Then, the wall part of the divergence of the velocity can be rewritten as follows:

$$\langle \nabla \cdot \mathbf{u} \rangle_{\mathrm{wall}\to i} = -\frac{d}{n^0} \sum_{j' \in \mathbb{V}_i} \left(\frac{n_j}{n_i}\right) \frac{\left\{\mathbf{u}_j - \left(-\mathbf{u}_i + 2\left[\mathbf{u}_j^{\mathrm{wall}} - \left(\mathbf{n}_j^{\mathrm{wall}} \cdot \mathbf{u}_j^{\mathrm{wall}}\right)\mathbf{n}_j^{\mathrm{wall}}\right]\right)\right\} \cdot \mathbf{r}_{ij'}}{\left\|\mathbf{r}_{ij'}\right\|^2} \omega_{ij'} . \tag{44}$$

Assuming that $\mathbf{u}_j^{\mathrm{wall}} \approx \mathbf{u}_i^{\mathrm{wall}}$, and $\mathbf{u}_{i'}$ given by Eq. (29), the Eq. (44) reads:

$$\langle \nabla \cdot \mathbf{u} \rangle_{\text{wall}\to i} = -\frac{d}{n^0} \sum_{j \in \mathbb{P}_{i'}} \left(\frac{n_j}{n_i}\right) \frac{(\mathbf{u}_j - \mathbf{u}_{i'}) \cdot (\mathbf{R}_i^{\text{ref}} \mathbf{r}_{i'j})}{\|\mathbf{r}_{i'j}\|^2} \omega_{i'j} \, . \tag{45}$$

### 2.2.6 Proposed technique to avoid misdetection of neighboring particles across zero-thickness walls

The adoption of zero-thickness polygon walls for thin shell FE can lead to incorrect interaction between particles placed at both sides of the wall, i.e., the misdetection of a neighboring particle $j$ inside the compact support of a particle $i$ even if the particles are separated by the wall. To overcome this problem, a simple but effective technique is proposed herein.

Since the wall is equidistant to a particle $i$ and its mirror $i'$, a potential neighboring particle $j$ belongs to the neighboring domain $\mathbb{P}_i$ when it is closer to $i$ than $i'$, i.e., the following criterion can be adopted:

$$\begin{cases} \text{If } \|\mathbf{r}_j - \mathbf{r}_i\| < \|\mathbf{r}_j - \mathbf{r}_{i'}\| & \rightarrow \quad j \in \mathbb{P}_i \\ \quad\quad \text{otherwise} & \rightarrow \quad j \notin \mathbb{P}_i \end{cases} . \tag{46}$$

To clarify the proposed criterion, Fig. 2-3 illustrates a generic situation in which the particle $j_1$ is regarded as a neighboring particle of $i$, while $j_2$ is not defined as a neighbor, although both are within the compact support of $i$.

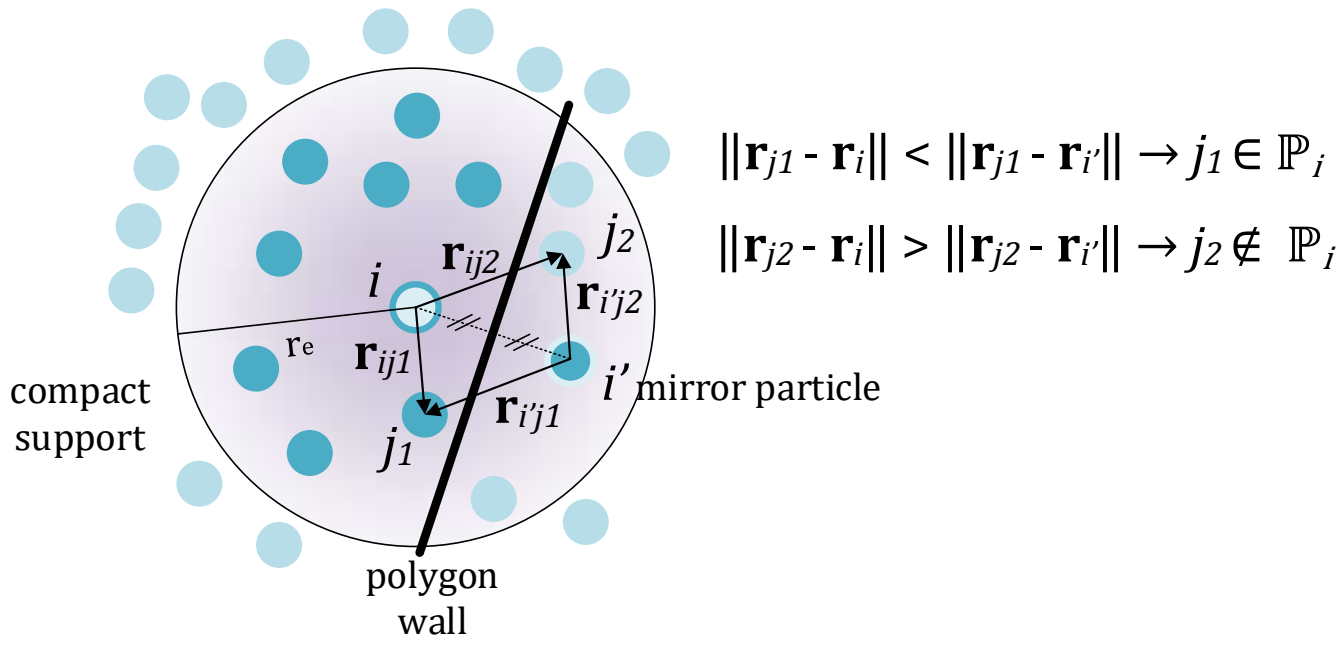

Fig. 2-3. Neighboring particle detection in the presence of zero-thickness walls.

Since the ERP model assumes that the polygon is a flat wall in the compact support, approximation errors occur when this assumption is not valid. In this way, the proposed technique may not work properly for curved boundary of relatively small radii of curvature inside the compact support. This topic concerns a spatial search issue, which is aim of future research.

### 2.2.7 Particle-mesh distance

Finally, aiming to speed up the search for fluid particles near polygons, an axis-aligned bounding box (AABB) hierarchy, implemented using the open source C++ library libigl [86], was used herein. The AABB is the rectangular box with the smallest volume within which all the vertices of the mesh lie, aligned with the axes of the coordinate system. The geometric data is converted into primitives in the AABB tree. From these primitives, a AABB hierarchy is created and used to speed up intersection and distance queries. If the mesh is static and positions of the points (particles) are updated, the initial AABB can be used for all the simulation. On the other hand, if the mesh geometry changes, the AABB needs to be update at each time step, which is a time-consuming task.

### 2.3 Solid domain

The description of this section refers to the geometrically exact shell theory, developed in a series of previous works, such as [87, 88, 89, 90]. In present work, only the fundamental aspects are discussed, for completeness of the presentation. For further details, the reader is invited to refer to here mentioned works.

#### 2.3.1 Kinematic description

The mid-surface of the shell is assumed to be planar at the initial (reference) configuration $r$. Let $\{\mathbf{e}_1^r; \mathbf{e}_2^r; \mathbf{e}_3^r\}$ be an orthonormal system, with corresponding coordinates $\{\xi_1; \xi_2; \zeta\}$, the vectors $\mathbf{e}_1^r$ and $\mathbf{e}_2^r$ are placed on the mid-surface plane and $\mathbf{e}_3^r$ is normal to this plane, as shown in Fig. 2-4.

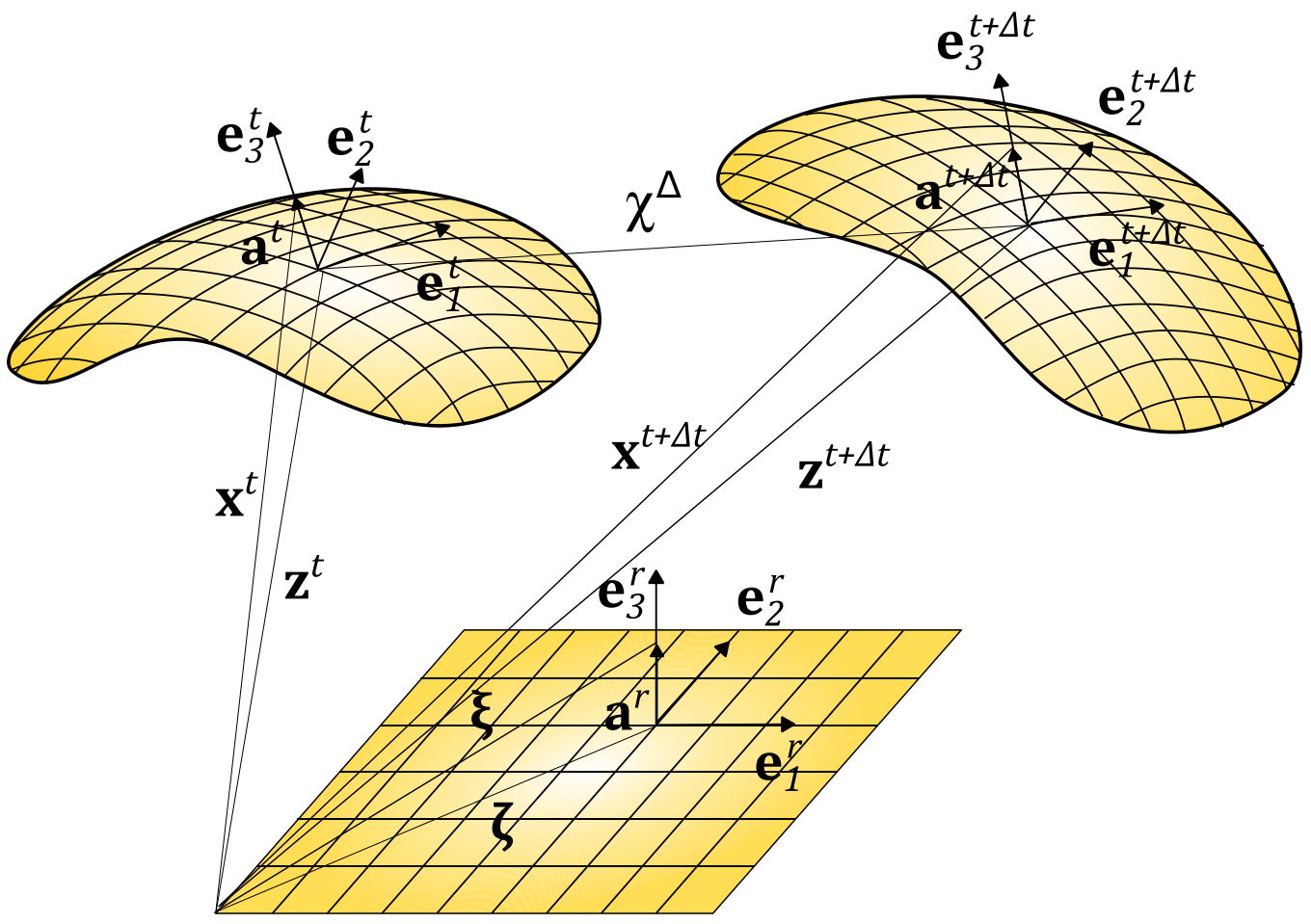

Fig. 2-4. Shell description and basic kinematical configurations through time evolution (adapted from Moreira [88]).

The position $\boldsymbol{\xi}$ of any shell material point in the reference configuration can be described by:

$$\boldsymbol{\xi} = \boldsymbol{\zeta} + \mathbf{a}^r, \tag{47}$$

where the vector $\boldsymbol{\zeta} = \xi_\alpha \mathbf{e}_\alpha^r$, $(\alpha = 1, 2)$, defines a material point on the reference mid-surface and $\mathbf{a}^r = \zeta \mathbf{e}_3^r$ is the shell director at this point, with $\zeta \in H = [-h^b, h^t]$ the thickness coordinate and $h = h^b + h^t$ being the shell thickness.

In the current configuration $t$ the position $\mathbf{x}$ of any material point can be represented by:

$$\mathbf{x}^t = \mathbf{z}^t + \mathbf{a}^t, \tag{48}$$

where $\mathbf{z}^t$ denotes the current position of a point in the mid-surface and $\mathbf{a}^t = \mathbf{Q}^t \mathbf{a}^r$ the current director at this point, with $\mathbf{Q}^t$ as the rotation tensor:

$$\mathbf{Q}^t = \mathbf{I} + \frac{4}{4 + \|\boldsymbol{\alpha}\|^2}\left(\mathbf{A} + \frac{1}{2}\mathbf{A}^2\right), \tag{49}$$

in which $\mathbf{A} = skew(\boldsymbol{\alpha})$ and $\boldsymbol{\alpha}$ is the Rodrigues rotation vector [91], given by:

$$\boldsymbol{\alpha} = \frac{\tan\left(\frac{\theta}{2}\right)}{\theta/2}\boldsymbol{\theta}, \tag{50}$$

where $\boldsymbol{\theta} = \theta \mathbf{e}$ is the Euler rotation vector representing an arbitrary finite rotation on 3D Euclidean space, with magnitude $\theta$ and axis direction given by the unit vector $\mathbf{e}$.

Similarly, the position of any material point at the end of the next instant, $t + \Delta t$, is given by:

$$\mathbf{x}^{t+\Delta t} = \mathbf{z}^{t+\Delta t} + \mathbf{a}^{t+\Delta t}. \tag{51}$$

Here $\mathbf{a}^{t+\Delta t} = \mathbf{Q}^\Delta \mathbf{a}^t$, where the superscript $\Delta$ stands for the quantities related to the interval $[t, t + \Delta t]$. Since $\mathbf{a}^t = \zeta \mathbf{e}_3^t$, Eq. (51) can be rewritten as:

$$\mathbf{x}^{t+\Delta t} = \mathbf{z}^{t+\Delta t} + \zeta \mathbf{Q}^\Delta \mathbf{e}_3^t\,. \tag{52}$$

The rotation tensor $\mathbf{Q}^\Delta$ stands for the incremental rotation from configuration $t$ to $t + \Delta t$ and is evaluated similarly to Eq. (49), but employing the incremental Rodrigues rotation vector.

### 2.3.2 Strain measures

Adopting the notation $(\cdot)_{,\alpha} = \partial(\cdot)/\partial\xi_\alpha$ for derivatives, the translational strain vector $\boldsymbol{\eta}_\alpha$ at instant $t + \Delta t$ is given by:

$$\boldsymbol{\eta}_\alpha^{t+\Delta t} = \mathbf{z}_{,\alpha}^{t+\Delta t} - \mathbf{e}_\alpha^{t+\Delta t}. \tag{53}$$

By multiplying both sides of Eq. (53) by $\mathbf{Q}^{t+\Delta t^T}$, its back-rotated counterpart is expressed by:

$$\boldsymbol{\eta}_\alpha^{t+\Delta t^r} = \mathbf{Q}^{t+\Delta t^T} \mathbf{z}_{,\alpha}^{t+\Delta t} - \mathbf{e}_\alpha^r. \tag{54}$$

Similarly, the back-rotated specific rotation vector $\boldsymbol{\kappa}_\alpha^r$ at instant $t + \Delta t$ can be expressed by:

$$\boldsymbol{\kappa}_\alpha^{t+\Delta t^r} = \mathbf{Q}^{t^T} \boldsymbol{\Xi}^{\Delta^T} \boldsymbol{\alpha}_{,\alpha}^\Delta + \boldsymbol{\kappa}_\alpha^{t^r}, \tag{55}$$

with $\boldsymbol{\Xi}^{\Delta}$ given by [91]:

$$\boldsymbol{\Xi}^{\Delta} = \frac{4}{4 + \|\boldsymbol{\alpha}^{\Delta}\|^2}\left(\mathbf{I} + \frac{1}{2}\mathbf{A}^{\Delta}\right), \tag{56}$$

where $\mathbf{A}^{\Delta}$ is the skew-symmetric tensor of $\boldsymbol{\alpha}^{\Delta}$ (the incremental Rodrigues rotation vector). With that, the generalized back-rotated strain vector $\boldsymbol{\varepsilon}^{t+\Delta t^r}$ can be represented as:

$$\boldsymbol{\varepsilon}^{t+\Delta t^r} = \begin{bmatrix} \boldsymbol{\varepsilon}_1^{t+\Delta t^r} \\ \boldsymbol{\varepsilon}_2^{t+\Delta t^r} \end{bmatrix}, \quad \text{with} \quad \boldsymbol{\varepsilon}_\alpha^{t+\Delta t^r} = \begin{bmatrix} \boldsymbol{\eta}_\alpha^{t+\Delta t^r} \\ \boldsymbol{\kappa}_\alpha^{t+\Delta t^r} \end{bmatrix}, (\alpha = 1,2). \tag{57}$$

### 2.3.3 Weak form of the equations of motion

As we establish here the equations of motion with the aid of Principle of Virtual Work (PVW), we start with the contributions from the internal ($\delta W_{int}$) and external ($\delta W_{ext}$) virtual works for the shell:

$$\delta W_{int} = \int_{\Omega_s} \left(\boldsymbol{\sigma}^{t+\Delta t^r} \cdot \delta\boldsymbol{\varepsilon}^{t+\Delta t^r}\right) d\Omega, \tag{58}$$

$$\delta W_{ext} = \int_{\Omega_s} (\bar{\mathbf{q}} \cdot \delta\mathbf{d}^{\Delta}) d\Omega, \tag{59}$$

where $\Omega_s$ is the shell reference mid-surface subdomain. The symbol $\delta$ represents virtual quantities. The generalized cross-sectional stress vector $\boldsymbol{\sigma}^{t+\Delta t^r}$ is calculated as:

$$\boldsymbol{\sigma}^{t+\Delta t^r} = \begin{bmatrix} \boldsymbol{\sigma}_1^{t+\Delta t^r} \\ \boldsymbol{\sigma}_2^{t+\Delta t^r} \end{bmatrix}, \quad \text{with} \quad \boldsymbol{\sigma}_\alpha^{t+\Delta t^r} = \begin{bmatrix} \mathfrak{n}_\alpha^{t+\Delta t^r} \\ \mathfrak{m}_\alpha^{t+\Delta t^r} \end{bmatrix}, (\alpha = 1,2), \tag{60}$$

in which $\mathfrak{n}_\alpha^{t+\Delta t^r}$ stands for the back-rotated cross-sectional forces and $\mathfrak{m}_\alpha^{t+\Delta t^r}$ is the back-rotated cross-sectional moments (both per unit length).

The generalized external forces vector $\bar{\mathbf{q}}$ is written as:

$$\bar{\mathbf{q}} = \begin{bmatrix} \bar{\mathbf{n}} \\ \boldsymbol{\Xi}^T \bar{\mathbf{m}} \end{bmatrix}, \tag{61}$$

where $\bar{\mathbf{n}}$ and $\bar{\mathbf{m}}$ are, respectively, vector of external forces and moments per unit reference area of the mid-surface.

The vector $\mathbf{d}^{\Delta}$ is defined as:

$$\mathbf{d}^{\Delta} = \begin{bmatrix} \boldsymbol{\chi}^{\Delta} \\ \boldsymbol{\alpha}^{\Delta} \end{bmatrix}, \tag{62}$$

where $\boldsymbol{\chi}^{\Delta}$ is the incremental displacement vector associated with any material point of the shell mid-surface.

Let $\delta T$ be the virtual work stemming from inertial effects, given by

$$\delta T = \delta T_1 + \delta T_2, \tag{63}$$

and $\delta T_1$ and $\delta T_2$ written as [90]:

$$\delta T_1 = \rho_s h \int_{\Omega_s} \ddot{\boldsymbol{\chi}}^{\Delta} \cdot \delta \boldsymbol{\chi}^{\Delta} d\Omega , \tag{64}$$

$$\delta T_2 = \frac{\rho_s h^3}{12} \int_{\Omega_s} \boldsymbol{\Xi}^T \left[ \mathbf{E}_3^{t+\Delta t^T} \mathbf{E}_3^{t+\Delta t} \dot{\boldsymbol{\omega}} + \mathbf{E}_3^{t+\Delta t^T} \left( \boldsymbol{\omega} \times \mathbf{E}_3^{t+\Delta t} \boldsymbol{\omega} \right) \right] \cdot \delta \boldsymbol{\alpha}^{\Delta} d\Omega \ , \tag{65}$$

where $\rho_s$ is the shell material specific mass, $\boldsymbol{\omega}$ designates the angular velocity vector, $\dot{\boldsymbol{\omega}}$ denotes the angular acceleration vector and $\mathbf{E}_3^{t+\Delta t} = skew(\mathbf{e}_3^{t+\Delta t})$. The vectors $\boldsymbol{\omega}$ and $\dot{\boldsymbol{\omega}}$ are the angular velocity and angular acceleration.

The principle of virtual work reads as:

$$\delta W_{int} - \delta W_{ext} + \delta T = 0, \tag{66}$$

for arbitrary incremental displacements $\delta \boldsymbol{\chi}^{\Delta}$ and incremental rotations $\delta \boldsymbol{\alpha}^{\Delta}$. This is the equation in which the finite element approximation is employed to compute the displacement and rotation fields.

The consistent linearization of Eq. (66) is necessary for establishing a nonlinear solution scheme, typically (and here) solved by the Newton-Raphson method. The reader can find derivation details in [87, 88, 90].

The material assumed for the shell structure is a hyperelastic model that, under small strain conditions, recovers linear elastic isotropic behavior, as presented in [87, 92].

The equations of motion for the structural model are integrated along time employing the Newmark (implicit) scheme, as detailed in [90].

### 2.3.4 Triangular shell finite element

The spatial discretization was done using the so-called t6-3i finite element [87], a six-node triangular element with quadratic shape functions for interpolation of the displacement field and linear shape functions for interpolation of the rotation field at the mid-points of the edges of the triangle. For a more detailed description of the shell modeling, the interested reader is referred to [87, 90].

## 2.4 Coupling scheme

The partitioned CSS scheme is adopted here, i.e., each subdomain is solved separately and sequentially at a given time level. At the beginning of the step, the fluid solver receives the structural positions and velocities and, afterwards, the solution of the fluid subdomain is updated. Then the hydrodynamic loads are sent to the structural nodes, and the structure responses are evaluated. Forces induced by the fluid particles are distributed on the FE nodes trough linear shape functions. Finally, nodal positions are updated, then imposing a new configuration of the wall boundary to the fluid solver. To handle different time scale between fluid and structural responses, sub-cycling technique might be adopted. Nevertheless, the same values of time steps for fluid and structure solvers are adopted herein to avoid additional approximations induced by the adoption of different time step intervals for each subdomain [45]. The proposed couple scheme can be summarized as below:

1. The force from a neighbor fluid particle $i \in \mathbb{P}_{\mathbb{E}_e}$ to each $e$-th finite element $\mathbb{E}_e$, $\mathbf{f}_{i\to\mathbb{E}_e}$, can be determined based on the normal and tangential components due to the pressure gradient and shear stress, respectively, at the fluid-structure interface, i.e., the reaction to the wall parts of the pressure gradient (Eq. (27)), viscous term (Eq. (28) or Eq. (29)) and repulsive force (Eq. (38)). Here, $\mathbb{P}_{\mathbb{E}_e}$ denotes the neighboring particles of a specific finite element $\mathbb{E}_e$, i.e., $\|\mathbf{r}_i - \mathbf{r}_i^{\mathbb{E}_e}\| < r_e$, with $\mathbf{r}_i^{\mathbb{E}_e} = \mathbf{r}_i^{\text{wall}}$ the closest point on the mesh to particle $i$.

$$\mathbf{f}_{i\to\mathbb{E}_e} = -\mathbf{f}_{\text{wall}\to i} = -(l_0)^d\left(-\langle\nabla P\rangle_{\text{wall}\to i} + \rho_f \nu_f \langle\nabla^2\mathbf{u}\rangle_{\text{wall}\to i} - \mathbf{f}_i^{\text{rep}}\right). \quad (67)$$

2. The contribution of the force $\mathbf{f}_{i\to\mathbb{E}_e}$ to the $j$-th vertices of the finite elements ($j = 1,2,3$), $\mathbf{f}_{i\to\mathbb{E}_{e,j}}$, is calculated by using linear shape functions $L_j^i = A_j^i/A$, related to the closest point $\mathbf{r}_i^{\mathbb{E}_e}$ on the finite element $\mathbb{E}_e$ to particle $i$:

$$\mathbf{f}_{i\to\mathbb{E}_{e,1}} = L_1^i\mathbf{f}_{i\to\mathbb{E}_e}, \quad \mathbf{f}_{i\to\mathbb{E}_{e,2}} = L_2^i\mathbf{f}_{i\to\mathbb{E}_e}, \quad \mathbf{f}_{i\to\mathbb{E}_{e,3}} = L_3^i\mathbf{f}_{i\to\mathbb{E}_e}. \quad (68)$$

   where $A$ is the total area of the element and $A_j$ stands for the subareas.

3. Since one vertex (node) can be shared by $m$ elements, the total force on each $j$-th vertex, is obtained by the following sum

$$\mathbf{f}_{\mathbb{P}_\mathbb{E}\to j} = \sum_{e\in m}\sum_{i\in\mathbb{P}_{\mathbb{E}_e}} \mathbf{f}_{i\to\mathbb{E}_{e,j}}, \quad (69)$$

   where $\mathbb{P}_\mathbb{E}$ represents all neighboring particles belonging to the $m$ elements $\mathbb{E}$ that share the vertex $j$..

4. Updated nodal positions are provided by the FE solver. Consequently, a new configuration of the wall boundary around the fluid particles is determined.

To evaluate the accuracy of the coupling scheme using ERP and linear shape functions, a quasi-static FSI benchmark test was simulated, and analytical and computed results were compared in Appendix A.

A summary of the implemented algorithm is illustrated in Fig. 2-5.

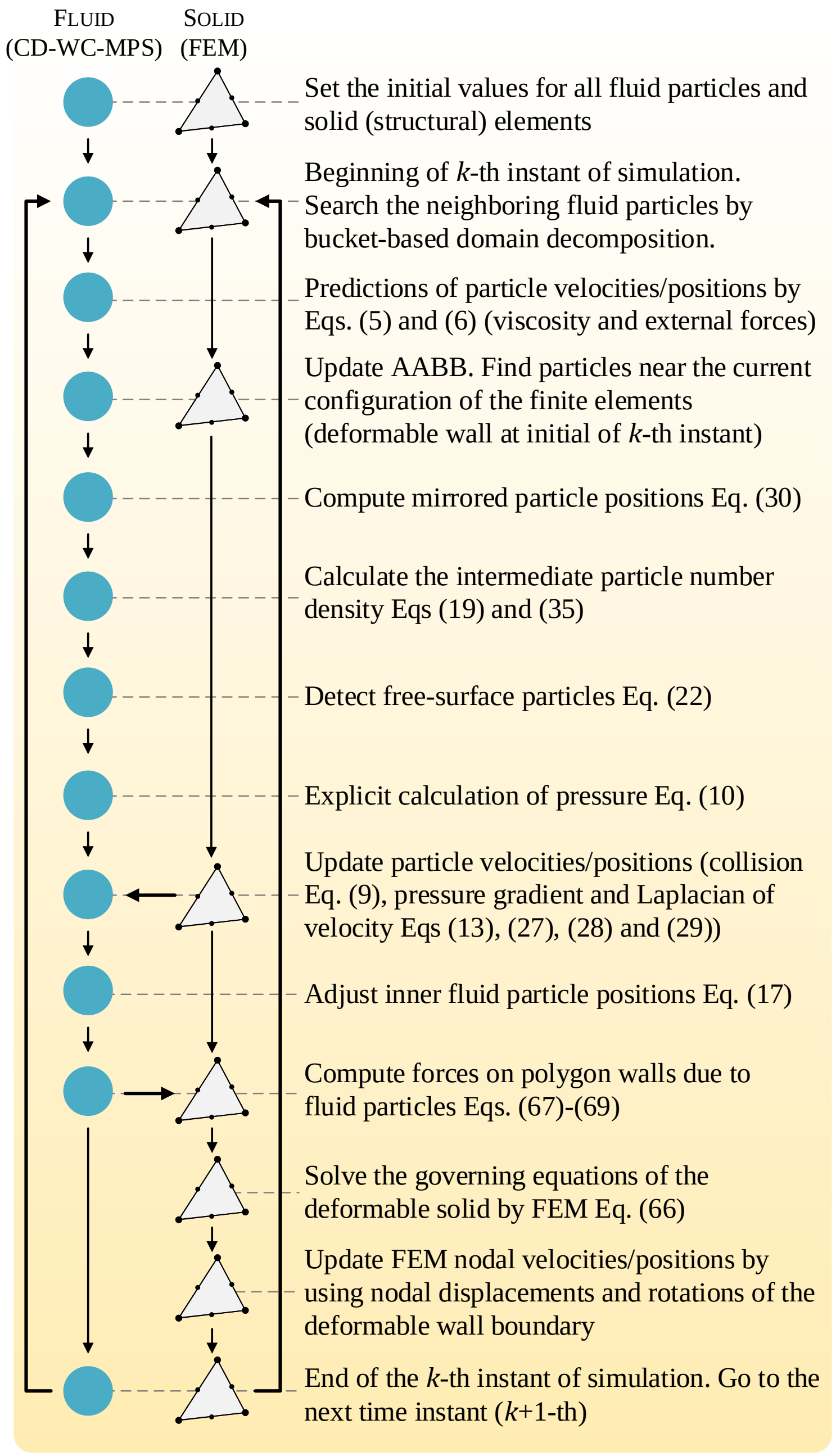

Fig. 2-5. Schematic diagram of the CD-WC-MPS coupled with geometrically nonlinear shell numerical algorithm (The reader interested on the bucket-based domain decomposition is referred to [93]).

It is important to point out that although the FE solver needs less CPU time than the CD-WC-MPS solver, one of the computational bottlenecks involving a high number of elements $\mathbb{E}$ is the construction of AABB at each time step (see item 4), which result in $\mathcal{O}(\mathbb{P}\log\mathbb{E})$ complexity, where ℙ is the number of particles. In this way, the finite element resolution was considered based on a compromise between accuracy and efficiency.

# 3 Numerical validation

## 3.1 WC-MPS benchmarks

For the validation of the improved WC-MPS proposed herein, two 3D benchmark tests were considered, namely hydrostatic tank and dam-break event. To illustrate the improvement on the pressure field by using the smoothed Continuity equation with Diffusive term, Eq. (19), here referred to as CD-WC-MPS, the computed results are compared against those obtained by adopting the Weight Function, Eq. (4), which is named as WF-WC-MPS model herein. No-slip boundary condition, Eq. (29) and Eq. (45), are considered. The fluid properties given in Table 3.1 and the gravity $g = 9.81\,\text{m/s}^2$ were considered for all simulations. The simulations were carried out using an Intel® Core™ Processor i7-4510U, processor base frequency of 2.00GHz, 4 MB cache, 4 cores and 8GB of RAM memory.

Table 3.1. Physical properties of the fluid.

| Property | Value |
|---|---|
| Density $\rho_f$ (kg/m³) | 1000 |
| Kinematic viscosity $\nu_f$ (m²/s) | $10^{-6}$ |

### 3.1.1 Hydrostatic pressure

This case consists of a rigid tank of height $H_T = 0.22$m and square bottom of side length $L_p = 0.2$m filled with water up to $H_F = 0.2$m, in hydrostatic condition, see Fig. 3-1. The simulation parameters are given in Table 3.2. Aimed to reach a static equilibrium, the dimensionless coefficient $A_F = 1$, see Eq. (14), was adopted, and three initial distances between particles, $l_0 = 20$, 10 and 5mm, were considered.

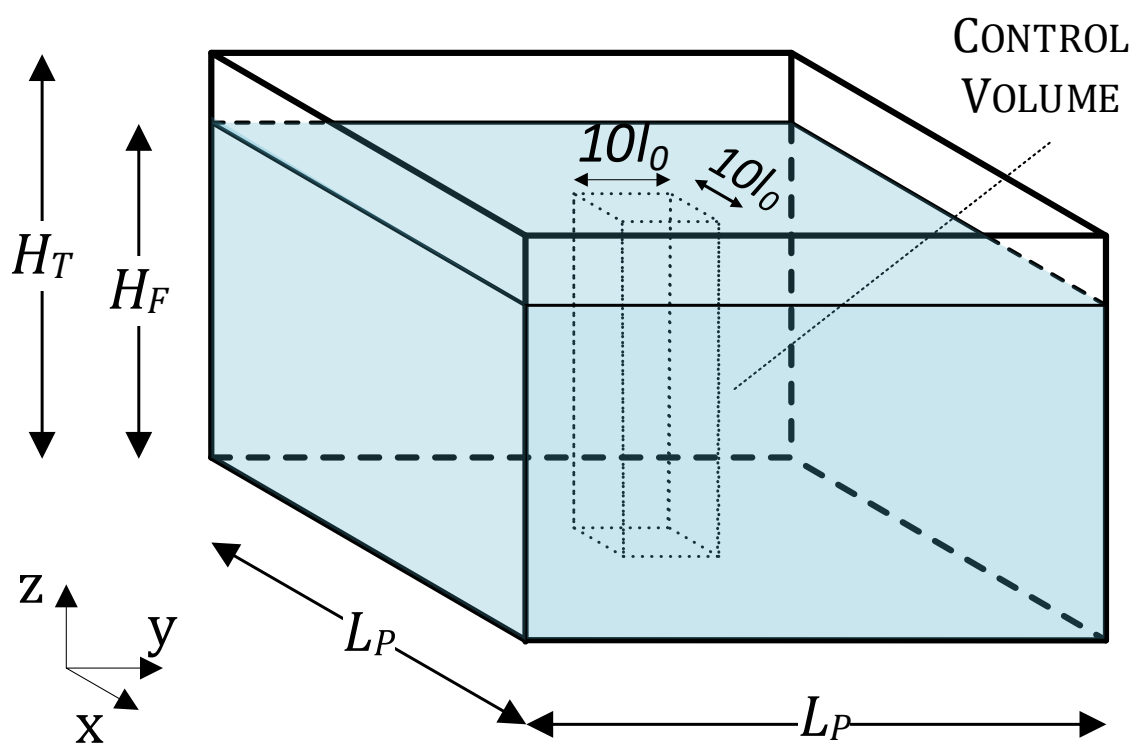

Fig. 3-1. Hydrostatic water column of height $H_F = 0.2$m in a rigid tank. Control volume of square cross section of side $10l_0$.

Table 3.2. Hydrostatic tank. Simulation parameters of the fluid.

| Parameter | Value | Parameter | Value |
|---|---|---|---|
| Particle distance $l_0$ (m) | 0.02, 0.01, 0.005 | Mach number $Ma$ | 0.1 |
| Time step $\Delta t_F$ (s) | 5.0, 2.5, 1.25 $\times 10^{-4}$ | Dimensionless number $A_F$ | 1 |
| Effective radius $r_e$ (m) | 2.1$\times l_0$ | Dimensionless constant $C_{rep}$ | 1 |
| Sound speed $c_0$ (m/s) | 15 | Surface thresholds ($\beta_{FS}, \varrho_{FS}$) | (0.98, 0.2) |
| Courant number $C_r$ | 0.2 | Simulation time (s) | 1.0 |

Fig. 3-2 depicts the analytical pressure steady solution and the non-dimensional pressure coefficient $C_P = P/\rho g H_F$ computed between the instants $t$ = 0.7 and 1.0s (quasi-static) for all particles within the control volume of square horizontal cross section of side $10 \times l_0$ and height $H_F$ (see Fig. 3-1). Due to some numerical inconsistencies on the approximated differential operators related to the truncation of the weight function at the free-surface particles close to the tank walls and the dynamic nature of the WC-MPS, variations of the positions of the particles and consequent spatial fluctuations of pressure field are expected even for still water tank. Despite out of the scope of the present work, the reader can find some works that have adopted different solutions (e.g., conservative pressure gradient operators, kernel type and smoothing length, higher-order numerical operators) to obtain an exact linear pressure field in the context of particle-based methods [73, 94, 95]. The pressure computed by CD-WC-MPS is remarkably improved, with much lower non-physical oscillations in relation to that computed by WF-WC-MPS. Moreover, the magnitude of the pressure dispersion is almost the same for all resolutions adopted. Hence, the accuracy of the proposed CD-WC-MPS is verified for all resolutions simulated in this quasi-static test.

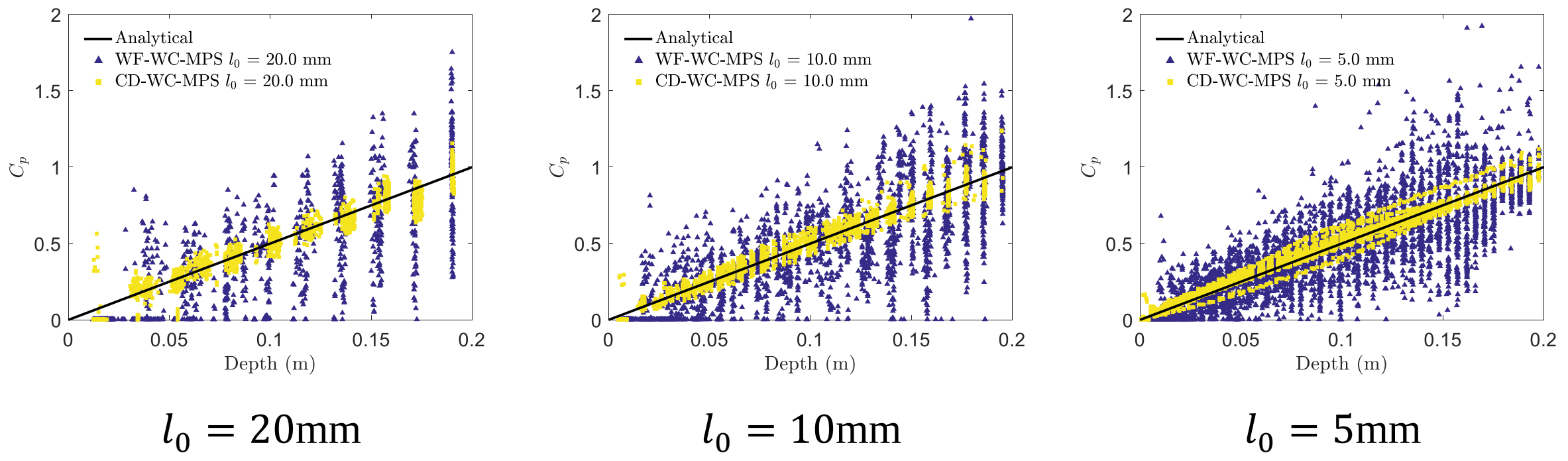

Fig. 3-2. Non-dimensional pressure ($C_p$) at the particles in the control volume (see Fig. 3-1) between $t$ = 0.7 and 1.0s. Comparison between the analytical and numerical results computed with WF-WC-MPS and CD-WC-MPS.

### 3.1.2 Dynamic pressure under dam-break event

Aiming to verify the improvements on the hydrodynamic pressure computation, the dam-break experiment conducted by Lobovský et al. [96] was simulated. Fig. 3-3 shows the initial configuration of the water column of height $H_F = 0.3$m and length $L_F = 0.6$m, rigid tank of length $L_T = 1.61$m, width $W_T = 0.15$m and height $H_T = 0.6$m, and the sensor S2 placed at the height $H_{S2} = 0.015$m. Five initial distances between particles, $l_0 = 20, 15, 10, 7.5$ and 5mm, were adopted to study the numerical convergence. The simulation parameters and computational times are shown in Table 3.3 and Table 3.4, respectively.

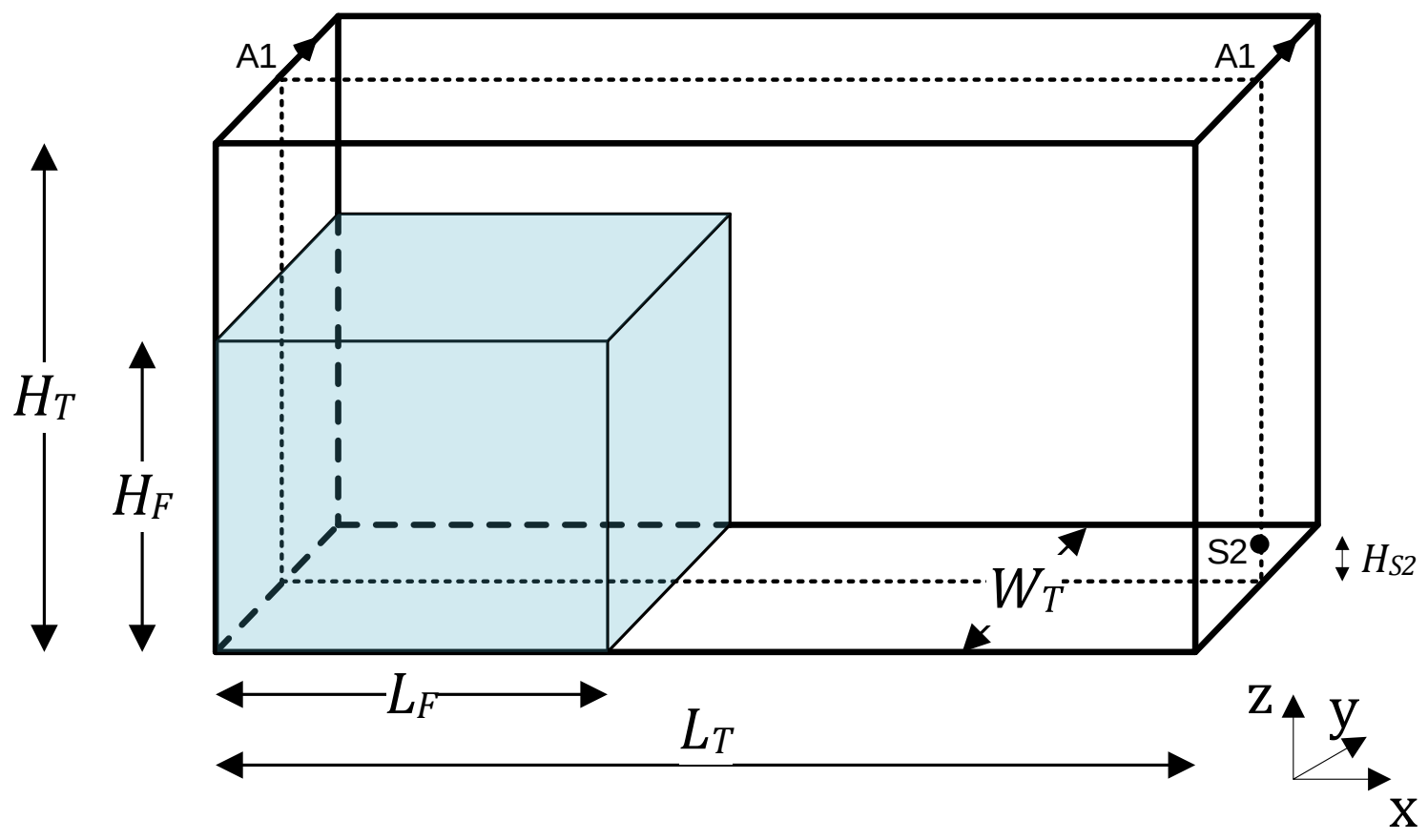

Fig. 3-3. Schematic drawing of the rigid tank, water column of height $H_F = 0.3$m and length $L_F = 0.6$m, and the sensor S2 placed at the height $H_{S2} = 0.015$m [96]. The section A1-A1 denotes the middle cross-sectional view.

Table 3.3. Dam break. Simulation parameters of the fluid.

| Parameter | Value | Parameter | Value |
|---|---|---|---|
| Particle distance $l_0$ (m) | 0.02, 0.015, 0.01, 0.075, 0.005 | Mach number $Ma$ | 0.1 |
| Time step $\Delta t_F$ (s) | 2.50, 2.50, 2.50, 1.50, 1.25 $\times 10^{-4}$ | Dimensionless number $A_F$ | 2 |
| Effective radius $r_e$ (m) | $2.1 \times l_0$ | Dimensionless constant $C_{rep}$ | 1 |
| Sound speed $c_0$ (m/s) | 15 | Surface thresholds ($\beta_{FS}, \varrho_{FS}$) | (0.98, 0.2) |

| | | |
|---|---|---|
| Courant number $C_r$ | 0.2 | |

Table 3.4. Dam break. Computational time.

| Particle distance (mm) | 20.0 | 15.0 | 10.0 | 7.5 | 5.0 |
|---|---|---|---|---|---|
| Simulation time (s) | | | 1.5 | | |
| Number of particles | 3150 | 8000 | 27000 | 64000 | 216000 |
| Time step (s) | $2.5\text{x}10^{-4}$ | $2.5\text{x}10^{-4}$ | $2.5\text{x}10^{-4}$ | $1.5\text{x}10^{-4}$ | $1.25\text{x}10^{-4}$ |
| Computation time | 0h28m | 1h00m | 3h40m | 8h00m | 48h00m |

Fig. 3-4 illustrates the present simulations using $l_0 = 5$mm of the collapsing water column at synchronized instants with the experiment. The color scale is associated to the pressure field of the middle cross-sectional view A1-A1 (see Fig. 3-3). The dimensionless time ($\tau$) is defined as $\tau = t\sqrt{g/H_F}$. The overall wave profile computed with the present WF-WC-MPS and CD-WC-MPS are in good agreement with the experiment. However, the simulation carried out using WF-WC-MPS leads to a rough pressure field. On the other hand, much smoother and continuous pressure field is obtained by applying the proposed CD-WC-MPS, resulting in a more stable and accurate calculation.

WF-WC-MPS | Experiment [96] | CD-WC-MPS

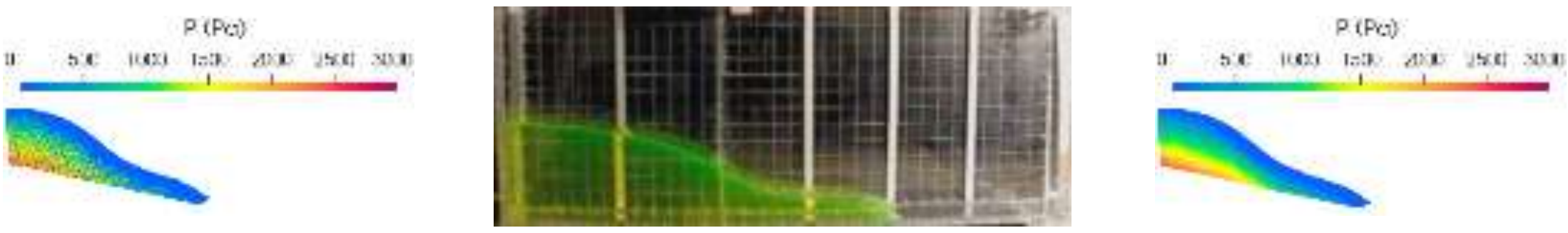

$\tau = 1.58$ ($t$ = 0.277 s)

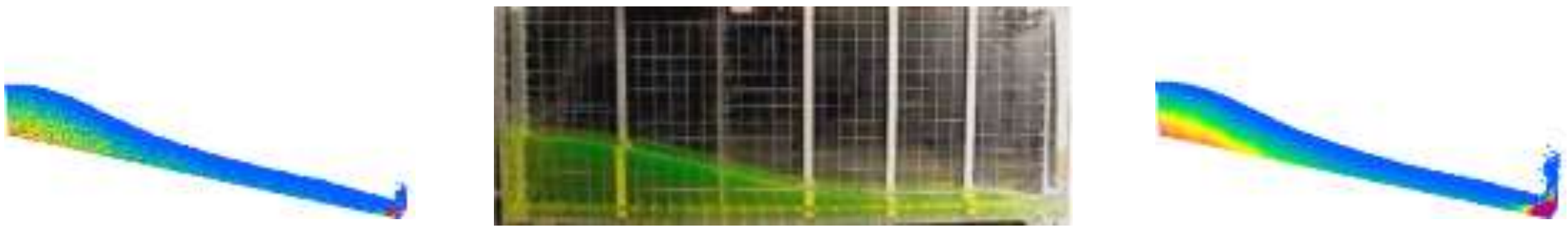

$\tau = 2.57$ ($t$ = 0.450 s)

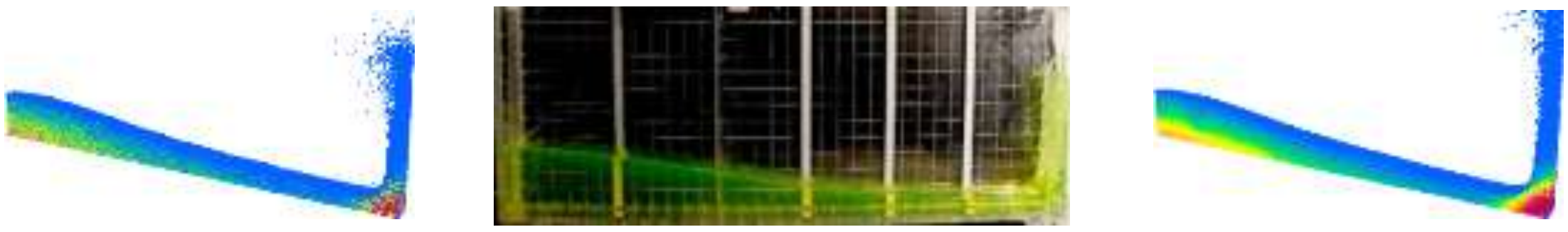

$\tau = 3.27$ ($t$ = 0.574 s)

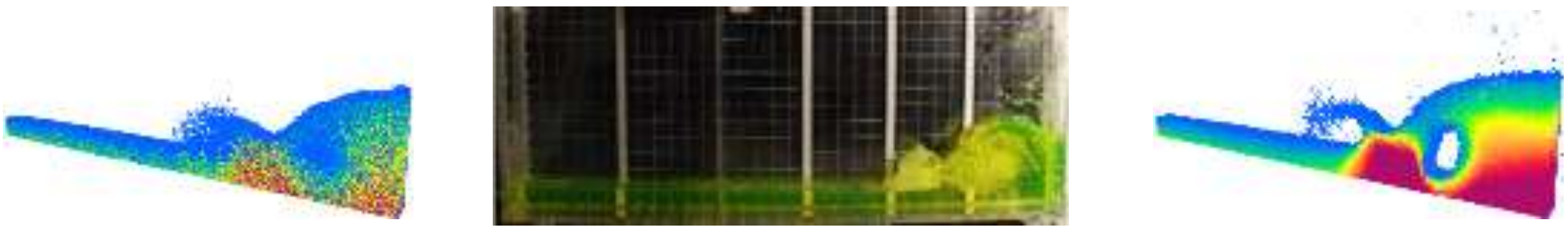

$\tau = 6.67$ ($t$ = 1.167 s)

Fig. 3-4. Dam break. Snapshots of the experiment [96] and numerical simulations (middle cross-sectional view A1-A1, see Fig. 3-3) at the instants $t$ = 0.277, 0.450, 0.574, 1.167s (non-dimensional times $\tau$ = 1.58, 2.57, 3.27, 6.67). WF-WC-MPS and CD-WC-MPS using the particle distance $l_0 = 5$mm. The colors on the fluid particles are related to its pressure magnitude.

In order to compare the numerical accuracy and dependency on the spatial resolution (convergence rate) of WF-WC-MPS and CD-WC-MPS schemes quantitatively, the $L_2$ norm error for the pressure, $L_2(p)$, at each discrete instant $m = 1, 2, \dots, M$ in the non-dimensional time interval $\tau \in [2.5, 7.0]$, is estimated as:

$$L_2(p) = \left( \sqrt{\frac{\sum_m^M \left(P_{num,m} - P_{exp,m}\right)^2}{\sum_m^M \left(P_{exp,m}\right)^2}} \right)_{2.5\leq\tau\leq7.0}, \qquad (70)$$

where $P_{num,m}$ and $P_{exp,m}$ are the numerically computed and experimentally measured pressures at sensor S2, respectively.

In Fig. 3-5(a), the simulations using CD-WC-MPS show a 1st order convergence rate, better than the convergence rate around 0.5 computed by WF-WC-MPS. Fig. 3-5(b) depicts the time histories of non-dimensional pressure coefficient $C_P = P/\rho g H_F$ measured in the experiment and numerically computed at sensor S2 using $l_0 = 5$ mm. The raw data of the pressure calculated at the fluid particle that is the closest one to the sensor position were considered without filtering or subsampling. A reasonable improvement on the stability and accuracy is obtained by CD-WC-MPS. One should note that between the instants $\tau$ = 2.5 and 5.6, and after $\tau$ = 6, approximately, the numerical results are slightly higher than the experimental ones. Such discrepancy, also observed in other particle-based simulations [97, 98, 73, 65], can be related to the non-physical expansion of volume due to the adoption particle shifting techniques, as recently investigated by Jandaguian et al. [99] and Lyu and Sun [100]. The adoption of recent enhanced particle collision/shifting models is considered as a future task. Despite of these overestimations, the adoption of the present CD-WC-MPS ensures stable and acceptable predictions of hydrodynamic problems as well as provides a better convergence rate.

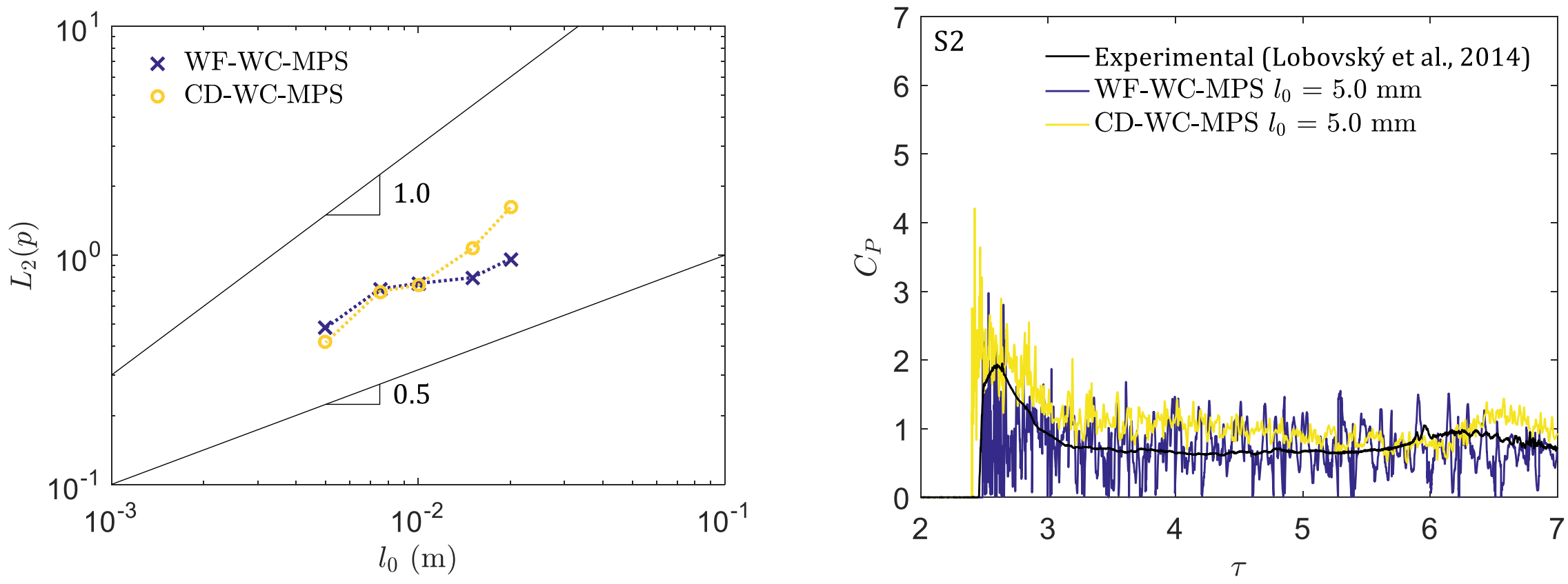

(a) $L_2$ norm error for pressure at Sensor S2 (b) Raw data of pressure at Sensor S2

Fig. 3-5. (a) $L_2(p)$ norm error for pressure as a function of the particle spacing $l_0$. (b) Raw data of experimental [96] and numerical pressures at sensor S2 using the particle distance $l_0 = 5$mm.

## 3.2 Validation of the FSI model

The validation of the proposed coupled WC-MPS-FE model were performed using two 3D dynamic benchmark experiments involving free-surface flow and thin-walled structures undergoing medium (sloshing) and large (dam breaking) deformations.

For all simulations, the proposed CD-WC-MPS model with no-slip boundary condition (Eq. (29) and Eq. (45)) and gravitational field $g = 9.81\text{m/s}^2$ were adopted. Deformable mesh with 16 elements per side were considered based on a compromise between accuracy of the structural model and efficiency of the spatial searching for the interaction of each element and surrounding particles.

### 3.2.1 Sloshing with an elastic plate

This case consists of a rectangular tank of length $L_T = 609$mm, width $W_T = 39$mm and height $H_T = 344.5$mm, partially filled with sunflower oil of initial height $H_F = 57.4$mm, and elastic plate of height Hp=57.4mm, width Wp=33.2mm and thickness $e_P = 4$mm, fixed at the midpoint ($L_P = 304.5$mm) of the tank bottom [101], as shown in Fig. 3-6(a). The tank is subjected to a harmonic rolling motion of amplitude 4 degrees and period $T = 1.646$s, see Fig. 3-6(b). Three initial distances between particles, $l_0 = 8, 4$ and 2.5mm ($\Delta x/l_0 = 1, 2$ and 3.2), were analyzed. $\Delta x$ means the average triangular element size. In addition, simulations using $l_0 = 4$mm combined with mesh sizes $\Delta x = 16, 8$ and 4mm were also simulated to verify the influence of the ratio $\Delta x/l_0$ on the numerical accuracy. Table 3.5 and Table 3.6 present respectively the physical properties of the fluid and elastic plate, and the simulation parameters. Table 3.7 gives the computational times.

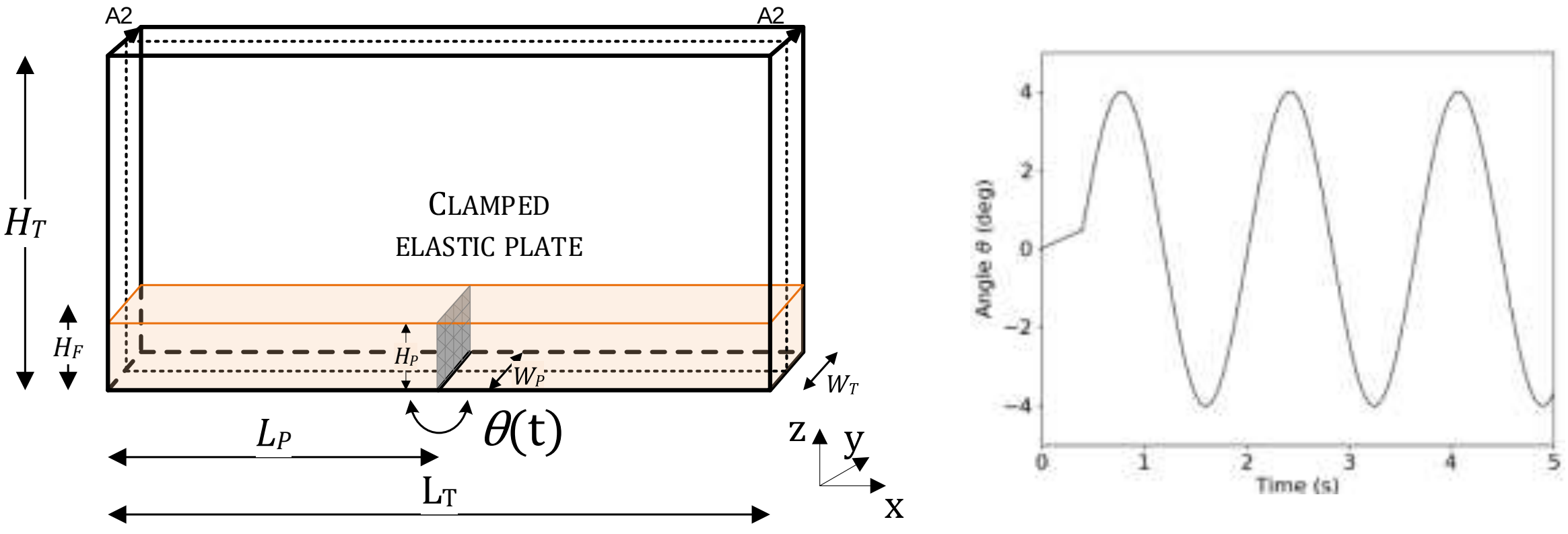

(a) (b)

Fig. 3-6. (a) Schematic drawing of the tank of length $L_T$ = 609mm, width $W_T$ = 39mm and height $H_T$ = 344.5mm, partially filled with sunflower oil of height $H_F$ = 57.4mm, and elastic plate of height $H_P$ = 57.4mm, width Wp=33.2mm and thickness $_P$ = 4mm, fixed at the midpoint ($L_P$ = 304.5mm). (b) Time history of the input motion [101]. The section A2-A2 represents the middle cross-sectional view

Table 3.5. Sloshing with an elastic plate. Physical properties of the sunflower oil and elastic plate.

| Fluid (sunflower oil) | | Elastic plate | |
|---|---|---|---|
| Property | Value | Property | Value |
| Density $\rho_f$ (kg/m³) | 917 | Density $\rho_S$ (kg/m³) | 1100 |
| Kinematic viscosity $\nu_f$ (m²/s) | 5.0×10$^{-5}$ | Young Modulus $E_s$ (MPa) | 6 |
| | | Poisson ratio $\nu_s$ | 0.49 |
| | | Thickness $e_p$ (m) | 0.004 |

Table 3.6. Sloshing with an elastic plate. Simulation parameters of the fluid and elastic plate.

| Fluid (sunflower oil) | | Elastic plate | |
|---|---|---|---|
| Parameter | Value | Parameter | Value |
| Particle distance $l_0$ (m) | 0.008, 0.004, 0.0025 | Element size $\Delta x$ (m) | 0.008 (0.016, 0.004) |
| Time step $\Delta t_F$ (s) | 2.5, 1.25, 1.0 ×10$^{-4}$ | Time step $\Delta t_S$ (s) | $\Delta t_F$ |
| Effective radius $r_e$ (m) | 2.1×$l_0$ | Mesh elements | 16×16 |
| Sound speed $c_0$ (m/s) | 15 | Rayleigh damping $\beta_R$ | 0.0001 |
| Courant number $C_r$ | 0.2 | Newmark coefficients ($\beta_N, \gamma_N$) [90] | (0.3, 0.5) |
| Mach number $Ma$ | 0.1 | | |
| Dimensionless number $A_F$ | 2 | | |
| Dimensionless constant $C_{rep}$ | 1 | | |
| Surface thresholds ($\beta_{FS}, \varrho_{FS}$) | (0.98, 0.2) | | |

Table 3.7. Sloshing with an elastic plate. Computational time.

| Particle distance (mm) | 8.0 | | 4.0 | | 2.5 |
|---|---|---|---|---|---|
| Simulation time (s) | | | 5.0 | | |
| Average element size (mm) | 8.0 | 16.0 | 8.0 | 4.0 | 8.0 |
| Number of particles | 2660 | | 21280 | | 89560 |
| Time step (s) | 2.5x10$^{-4}$ | | 1.25x10$^{-4}$ | | 1x10$^{-4}$ |
| Computation time | 6h30m | 9h20m | 15h50m | 40h00m | 39h30m |

Fig. 3-7 shows the comparison between experimental and numerical wave evolution and plate deformation. The color scale of the fluid particles is related to its pressure field. From the snapshots, the numerical calculations provide a smooth pressure field and are in very good agreement with the experimental one. In general, the proposed model is able to well reproduce the main hydrodynamic characteristics of the free-surface flow as well as the structural dynamic behavior of the elastic plate.

Fig. 3-8(a) shows the horizontal displacement of the elastic plate tip, relative to the local reference frame fixed to the tank, obtained by numerical simulation and compared against the experimental and numerical results using particle finite element method (PFEM) [101]. The overall trends of the computed displacements agree well with the experimental one and performed slightly better than PFEM results, although small discrepancy occurs near $t = 0.5$s and the experimental motion is not numerically well reproduced. This discrepancy was also found in the numerical results from others [102 7 103 12 45 59]. These differences can be related to some uncertainties in the experiments. According to Botia Vera [104], repeatable and symmetry were not completely achieved in this shallow sloshing experiment, possible due to some pre deformation of the beam in conjunction with some material hysteresis. The comparisons show that as the spatial resolution increases the WC-MPS-FE results converge to the experimental data.

Fig. 3-8(b) shows that the proposed coupled WC-MPS-FE model provides accurate results for distinct particle-mesh sizes even for a large ratio of $\Delta x/l_0 \approx 4$, which significantly increases the computational time efficiency with the reduced complexity of the particle-mesh search based on AABB.

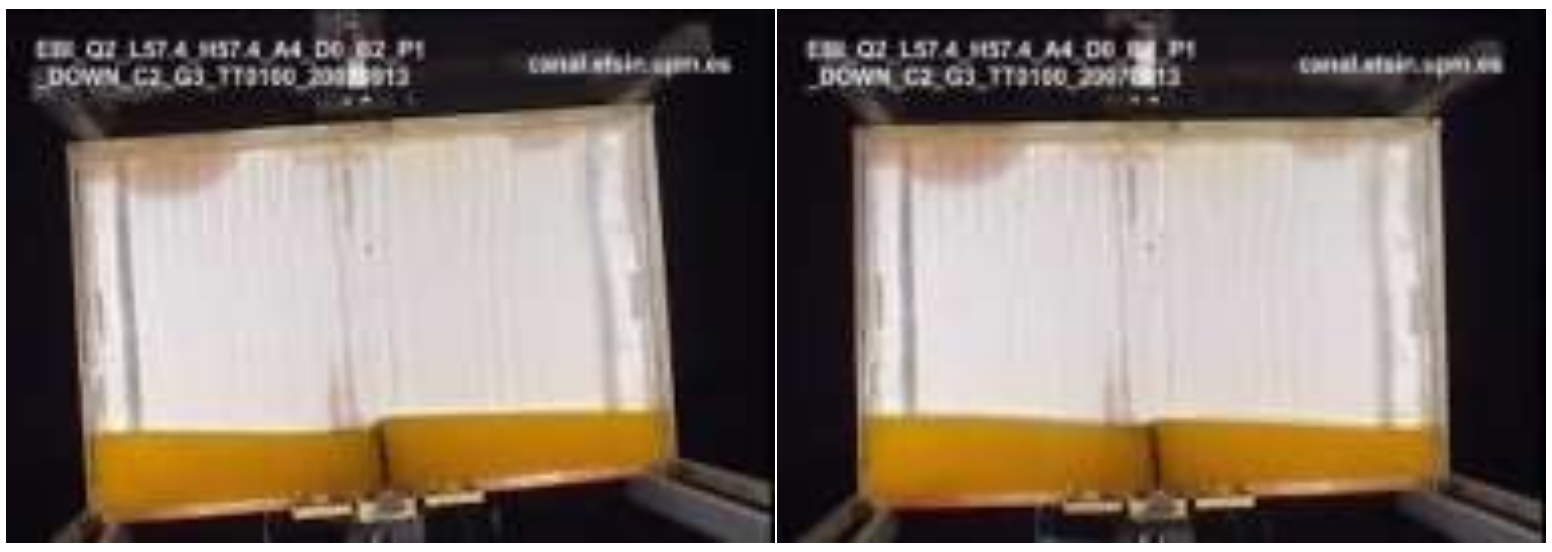

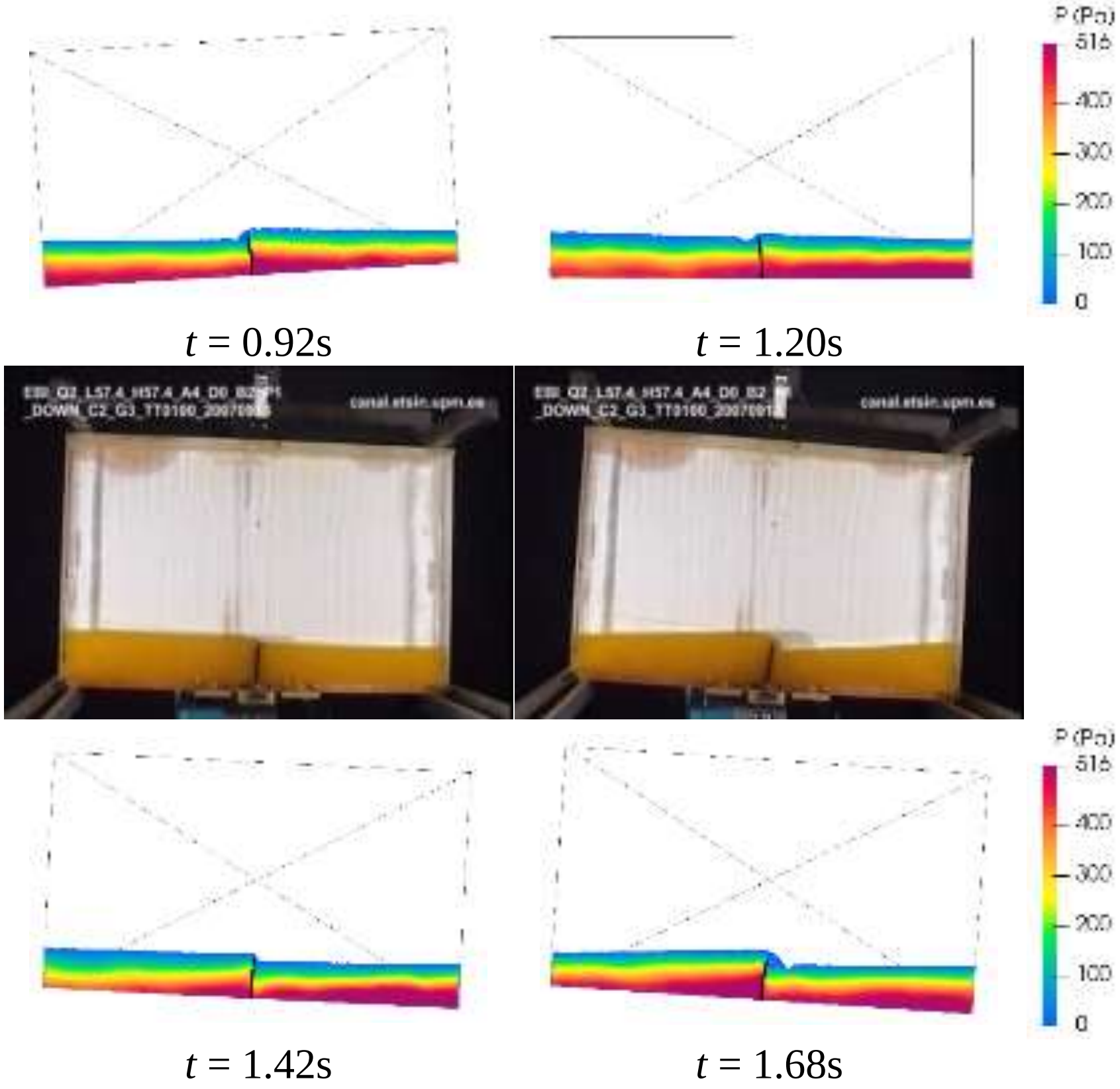

Fig. 3-7. Sloshing with a clamped gate. Snapshots of the experiment [101] and numerical simulations (middle cross-sectional view A2-A2, see Fig. 3-6) with WC-MPS-FE for the particle distance $l_0 = 2.5$mm at the instants $t$ = 0.92, 1.20, 1.42, 1.68s. The colors on the fluid particles are related to its pressure magnitude

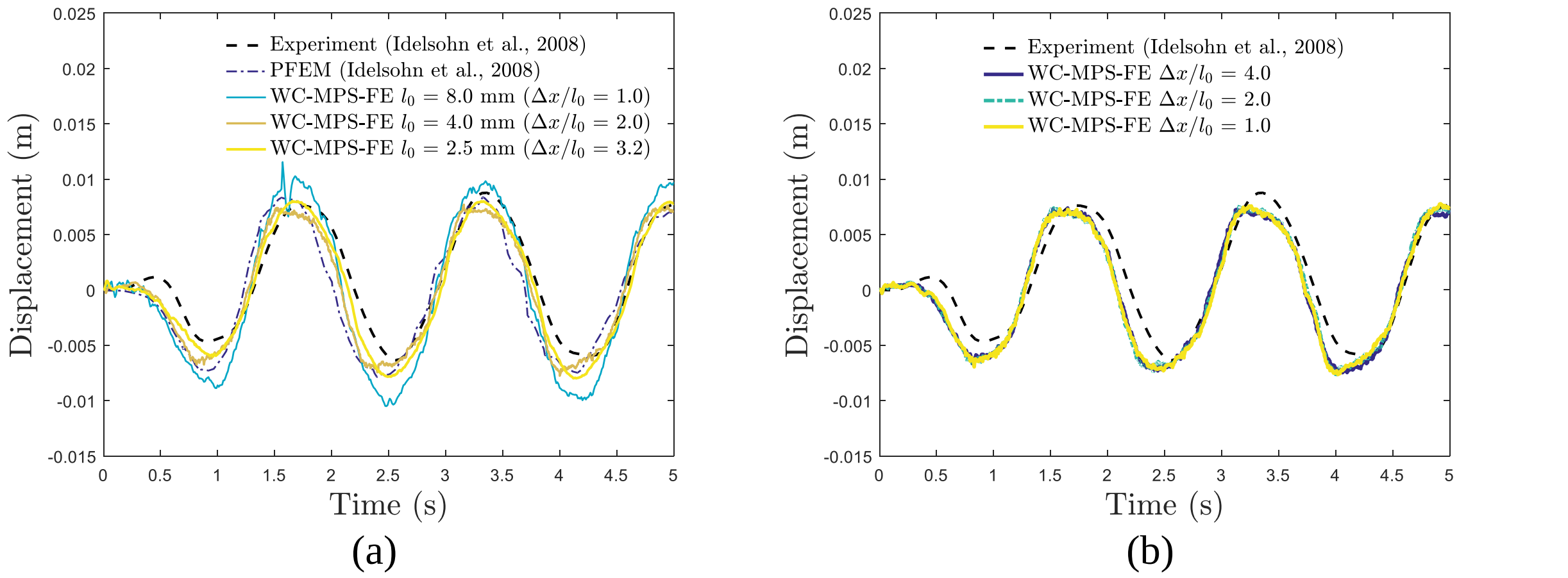

Fig. 3-8. Time series of the horizontal displacements of the free end of the clamped plate. (a) Experimental data and numerical results of PFEM from Idelsohn et al. [101] and present WC-MPS-FE with $l_0 = 8, 4$ and 2.5mm and $\Delta x = 8$mm. (b) Experimental data and present WC-MPS-FE with $l_0 = 4$mm and $\Delta x = 16, 8$ and 4mm ($\Delta x / l_0 = 4, 2$ and 1).

### 3.2.2 Dam breaking hitting an elastic plate clamped at one edge

In the last case, the proposed coupled model is applied to predict the violent transient free-surface flow interacting with elastic structures undergoing large deformations. Fig. 3-9 shows the initial configuration of the experiment conducted by Liao et al. [105], consists of a tank of length $L_T = 0.8$ m, width $W_T = 0.2$ m and height $H_T = 0.6$ m, a water column of height $H_F = 0.4$ m, length $L_F = 0.2$ m and confined by a rigid gate. An elastic plate of height $H_P = 0.1$m, thickness $e_p = 0.004$m and width $W_P = 0.1995$m is clamped at the bottom and located $L_P = 0.6$ m downstream. Three distances between particles, $l_0 = 20, 10$ and $5$ mm ($\Delta x/l_0 = 1.35, 2.7$ and $5.4$), were adopted in the simulations. The influence of $\Delta x/l_0$ on the numerical accuracy varying $\Delta x = 54, 27$ and $13.5$mm was analyzed considering $l_0 = 10$mm. The physical properties of the fluid and elastic plate can be found in Table 3.8 and the simulation parameters are given in Table 3.9. The computational times are presented in Table 3.10.

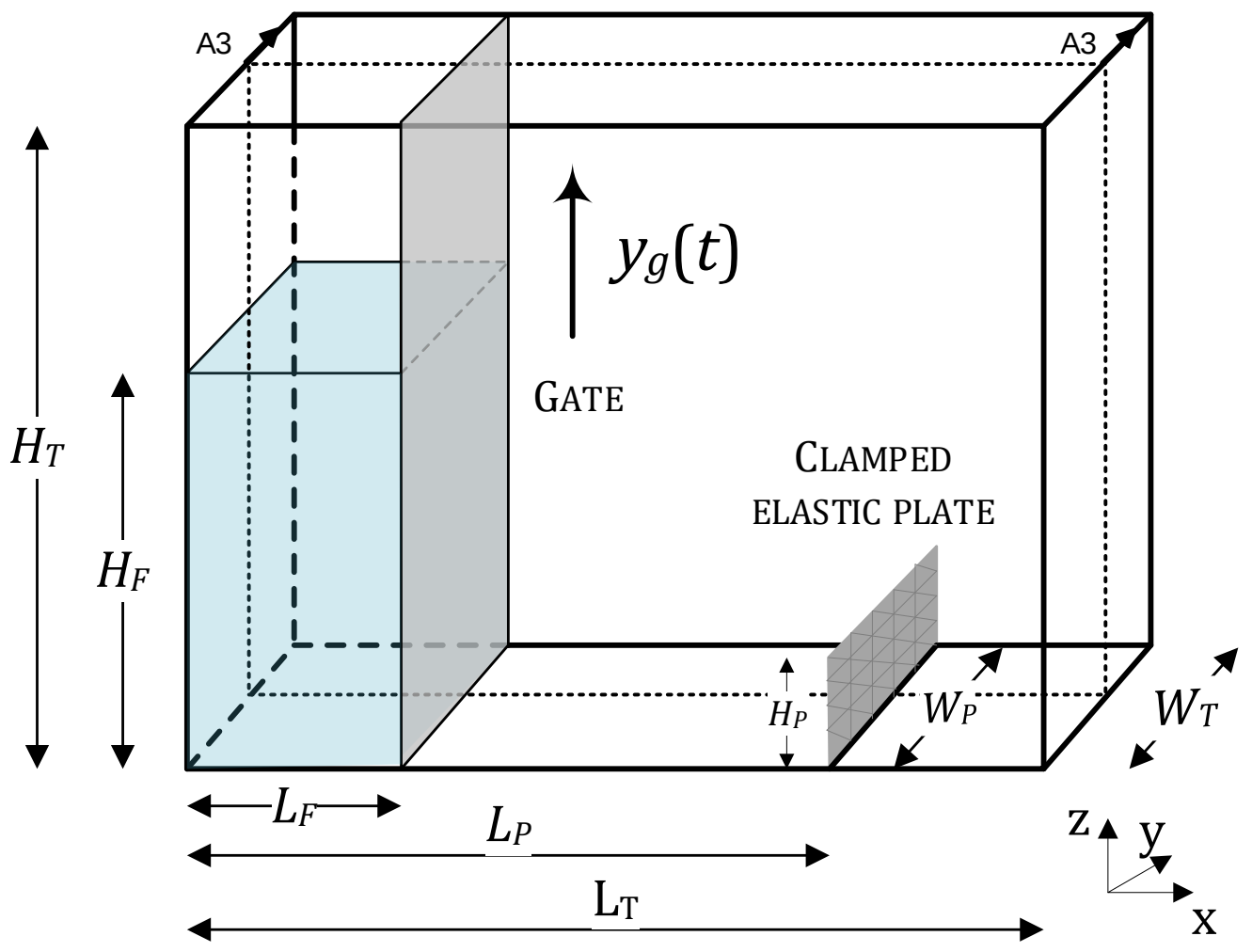

Fig. 3-9. Schematic drawing of the tank of length $L_T = 0.8$m, width $W_T = 0.2$m and height $H_T = 0.6$m, water column of height $H_F = 0.4$m and length $L_F = 0.2$m, clamped elastic plate of height $H_P = 0.1$m, width $W_P = 0.1995$m and thickness $e = 0.002$m, and located $L_P = 0.6$m downstream. The gate is subjected to an upward motion $y_g(t)$ [105]. The section A3-A3 represents the middle cross-sectional view.

In the present simulations, the vertical motion of the gate follows a smooth function $y_g(t)$ that approximately fits the experimental vertical motion provided in Liao et al. [105]:

$$y_g(t) = -300t^3 + 75t^2 \,. \tag{71}$$

Table 3.8. Dam breaking hitting a clamped plate. Physical properties of the fluid and elastic plate.

| Fluid | | Elastic plate | |
|---|---|---|---|
| Property | Value | Property | Value |
| Density $\rho_f$ (kg/m³) | 1000 | Density $\rho_S$ (kg/m³) | 1161.54 |
| Kinematic viscosity $\nu_f$ (m²/s) | $10^{-6}$ | Young Modulus $E_S$ (MPa) | 3.5 |
| | | Poisson ratio $\nu_s$ | 0.3 |
| | | Thickness $e_p$ (m) | 0.004 |

Table 3.9. Dam breaking hitting a clamped plate. Simulation parameters of the fluid and elastic plate.

| Fluid | | Elastic plate | |
|---|---|---|---|
| Parameter | Value | Parameter | Value |
| Particle distance $l_0$ (m) | 0.02, 0.01, 0.005 | Element size $\Delta x$ (m) | 0.027 (0.054, 0.0135) |
| Time step $\Delta t_F$ (s) | 5, 2.5, 1.0 $\times 10^{-4}$ | Time step $\Delta t_S$ (s) | $\Delta t_F$ |
| Effective radius $r_e$ (m) | $2.1 \times l_0$ | Mesh elements | 16×16 |
| Sound speed $c_0$ (m/s) | 15 | Rayleigh damping $\beta_R$ | 0.001 |
| Courant number $C_r$ | 0.2 | Newmark coefficients ($\beta_N, \gamma_N$) [90] | (0.3, 0.5) |
| Mach number $Ma$ | 0.1 | | |
| Dimensionless number $A_F$ | 2 | | |
| Dimensionless constant $C_{rep}$ | 1 | | |
| Surface thresholds ($\beta_{FS}, \varrho_{FS}$) | (0.98, 0.2) | | |

Table 3.10. Dam breaking hitting a clamped plate. Computational time.

| Particle distance (mm) | 20.0 | | 10.0 | | 5.0 |
|---|---|---|---|---|---|
| Simulation time (s) | | | 1.0 | | |
| Average element size (mm) | 27.0 | 54.0 | 27.0 | 13.5 | 27.0 |
| Number of particles | 2000 | | 16000 | | 128000 |
| Time step (s) | $5\text{x}10^{-4}$ | | $2.5\text{x}10^{-4}$ | | $1.0\text{x}10^{-4}$ |
| Computation time | 0h35m | 1h45m | 3h50m | 7h30m | 24h00m |

Fig. 3-10 shows the sequences of the flow, at selected instants computed using particle distance $l_0 = 5\text{mm}$ ($\Delta x / l_0 = 5.4$). The colors scale of the fluid particles denotes the non-dimensional velocity $v/(2\sqrt{gH_F})$ and the color scale on the elastic plate refers to its displacement. As the fluid is released, the dam-break flow proceeds and the wave front hits the clamped plate at an instant just before to $t = 0.25\text{s}$. After that, the plate undergoes a large displacement and part of the fluid is deflected upward at $t = 0.35\text{s}$. Subsequently, at $t = 0.45\text{s}$, the wave impact on the downstream wall generates a vertical run-up jet whereas a back flow of part of the fluid propagates along the tank bottom toward the largely deformed elastic plate. Afterwards, at the instants $t = 0.60\text{s}$ and $t = 0.80\text{s}$, the splashed fluid falls due to the gravity, followed by a merging with the back flow near the elastic plate, creating a violent turbulent cavity flow. Concerning the elastic plate, a second impact on its right-side lead to a reverse deflection.

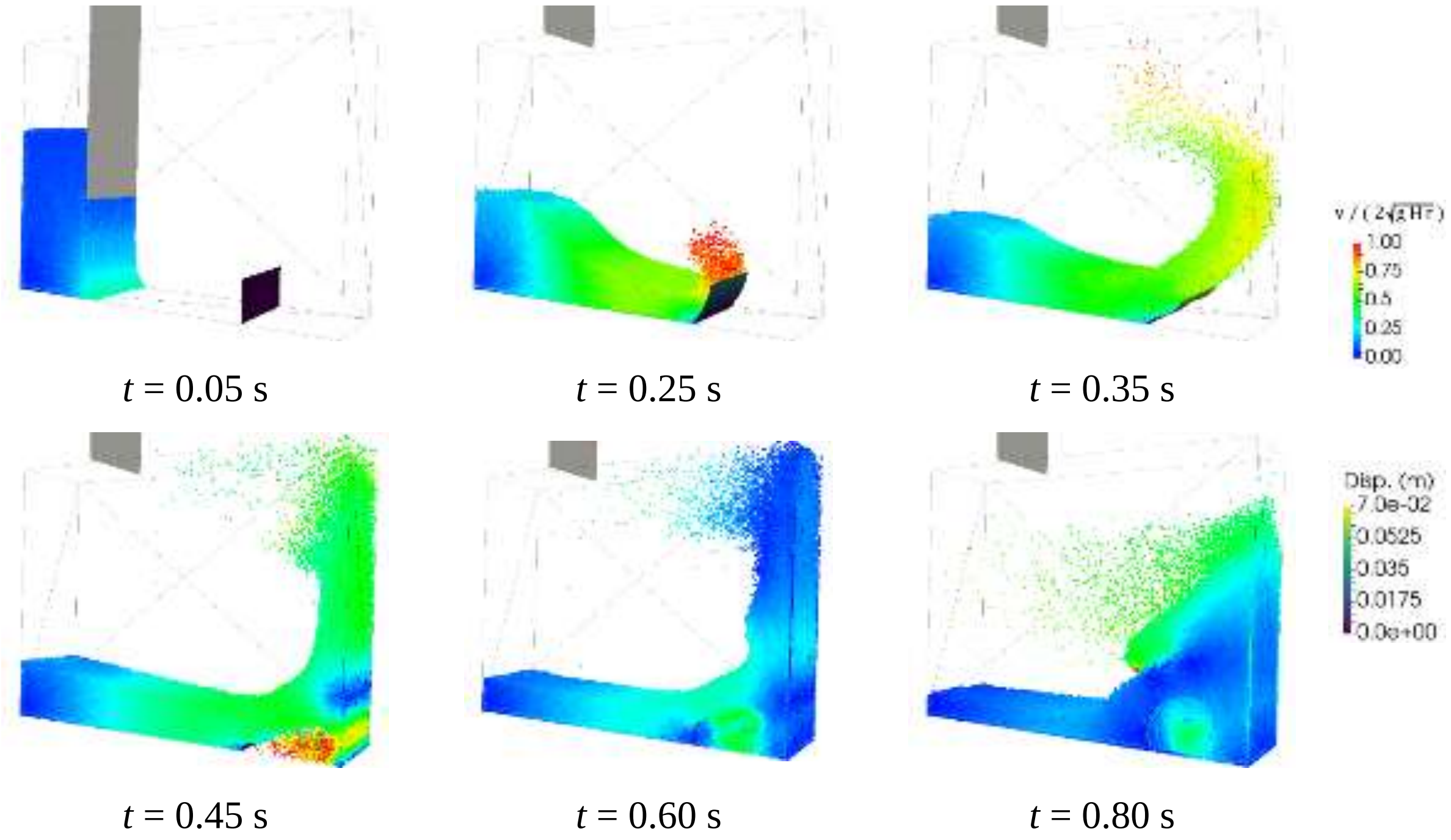

Fig. 3-10. Dam breaking hitting a clamped plate. Snapshots of the numerical simulations with WC-MPS-FE for the particle distance $l_0 = 5$mm at the instants $t$ = 0.05, 0.25, 0.35, 0.45, 0.60, 0.80s. The colors on the particles are related to its non-dimensional velocity $v/(2\sqrt{gH_F})$ and on the elastic plate are related to its displacement.

Free-surface profile and plate deformation at the instants $t$ = 0.25, 0.28, 0.30, 0.55, 0.60 and 0.80s from the experiment [105] and the simulation carried in the present study using the proposed WC-MPS-FE with particle distance $l_0 = 5$mm are compared in Fig. 3-11. The colors scale of the fluid particles denotes the non-dimensional velocity $v/(2\sqrt{gH_F})$ and the color scale of the elastic plate designates its displacement, both shown partially the middle cross-sectional view A3-A3 (see Fig. 3-9). During the initial stages $0.25 \leq t \leq 0.3s$, the wave front hits the elastic plate and subsequently overtops it towards the downstream tank wall. Advancing in time, the back flow hits the right-side of the plate between the instants $t =$ 0.55 and 0.60s. As a result, an open cavity surrounded by the fluid is created, which increases the local vortex intensity, followed by considerable plate deformation. Afterwards, at the instant $t = 0.80$s, part of collapsed runup fluid merges with the back flow near the elastic plate, violent rotational flow and the open cavity still remain, and the plate is deflected in the reverse direction. Up to $t = 0.30$s, the computed local free-surface profiles and plate deformations agree very well with the experimental ones. Despite some differences from $t = 0.55$ s to $t = 0.80$ s, the main behavior of this complex interaction is numerically reproduced. The main reasons of these differences might be the neglection of the air-phase in the present simulations.

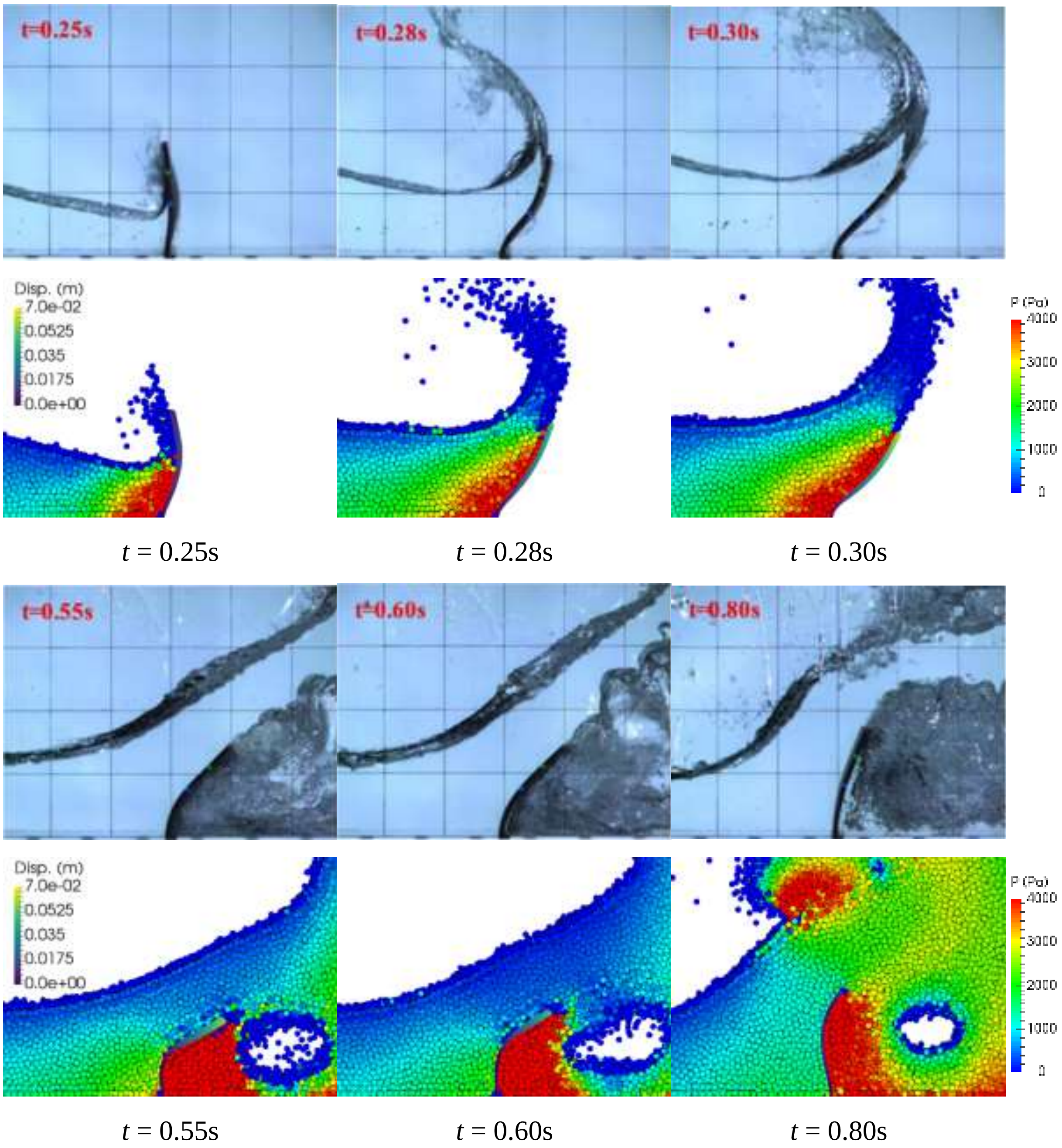

Fig. 3-11. Dam breaking hitting a clamped plate. Snapshots of the experiment [105] and numerical simulations (middle cross-sectional view A3-A3, see Fig. 3-9) with WC-MPS-FE for the particle distance $l_0 = 5$mm at the instants *t* = 0.25, 0.28, 0.30, 0.55, 0.60, 0.80s. The colors on the fluid particles are related to its pressure magnitude.

Fig. 3-12(a) provides the time-histories of horizontal displacement of the elastic plate tip measured experimentally [105] and computed numerically with single-phase and multi-phase SPH [23], single-phase MPS [22], and present single-phase WC-MPS-FE using different initial particle distances $l_0 = 20, 10$ and $5$mm ($\Delta x / l_0 = 1.35, 2.7$ and $5.4$). The wave front hits the bottom of the plate leading to a negative acceleration of the plate's tip, which is evidenced by the negative displacement approximately at $t = 0.25$s. Despite both experimental and numerical displacements present the same initial trend, the experimental impact occurs at approximately $t$ = 0.27s while the numerical ones, computed by the present model, happen around $t$ = 0.25s. This discrepancy was also observed in Liao et al. [105] and attributed to the neglection of the gate motion in the numerical simulation. However, since the influence of the gate is numerically reproduced herein, this reason can be disregarded here. A possible reason for this discrepancy might be attributed to the Laplacian operator of accuracy order $\mathcal{O}(l_0^{-1})$ 106, 107 used for the wall contribution (Eq. (29)) or even the model resolution. Nevertheless, in order to rigorously clarify such discrepancy further investigations should be considered. After the first wave impact, the displacement at the plate's tip suddenly increases. Between the instants $t$ = 0.4 and 0.6s, a small oscillation in the displacement was experimentally measured. However, such oscillation was not computed with the present model. As pointed by Liao et al. [105], higher modes of structural vibration are caused by the presence of entrapped air near the elastic plate so that the numerical modeling of the air-phase is required to reproduce this physical phenomenon appropriately. Others results numerically computed with single-phase and multi-phase particle methods [22, 23] are plotted in Fig. 3-12 and illustrate the importance of the air-phase modeling at this time interval.

After the instant $t$ = 0.5s, the displacement of the plate is gradually reduced due to the impact of the back flow. Besides small discrepancies, the overall trend of the displacement experimentally measured by Liao et al. [105] is correctly reproduced by the proposed model, even when employing a coarse resolution on the fluid description.

Fig. 3-12(b) shows that there is no noticeable difference in the computed horizontal displacements from the proposed WC-MPS-FEM using $\Delta x = 27$ or 13.5mm, i.e., $\Delta x / l_0 = 1.35, 2.7$. However, when using $\Delta x = 54$mm, i.e., large ratio $\Delta x / l_0 = 5.4$, discrepancy of the computed displacements occurs. Since Fig. 3-12(a) shows that the displacements using small particle distance $l_0 = 5$mm, with the same ratio $\Delta x / l_0 = 5.4$, are in reasonable agreement with the reference experimental and numerical solutions, this indicates the feasibility of the proposed WC-MPS-FEM to deal with large deformations using a high ratio $\Delta x / l_0$ when a fine particle resolution is adopted. Table 3.10 illustrates how the number of finite elements, keeping the particle size, affects the computational performance in the cases with deformable mesh, since the AABB tree needs to be updated in every calculation step. This help us to demonstrate how important is the feature of the present model in provide reasonable accurate results for large ratios between mesh size and particle distance.

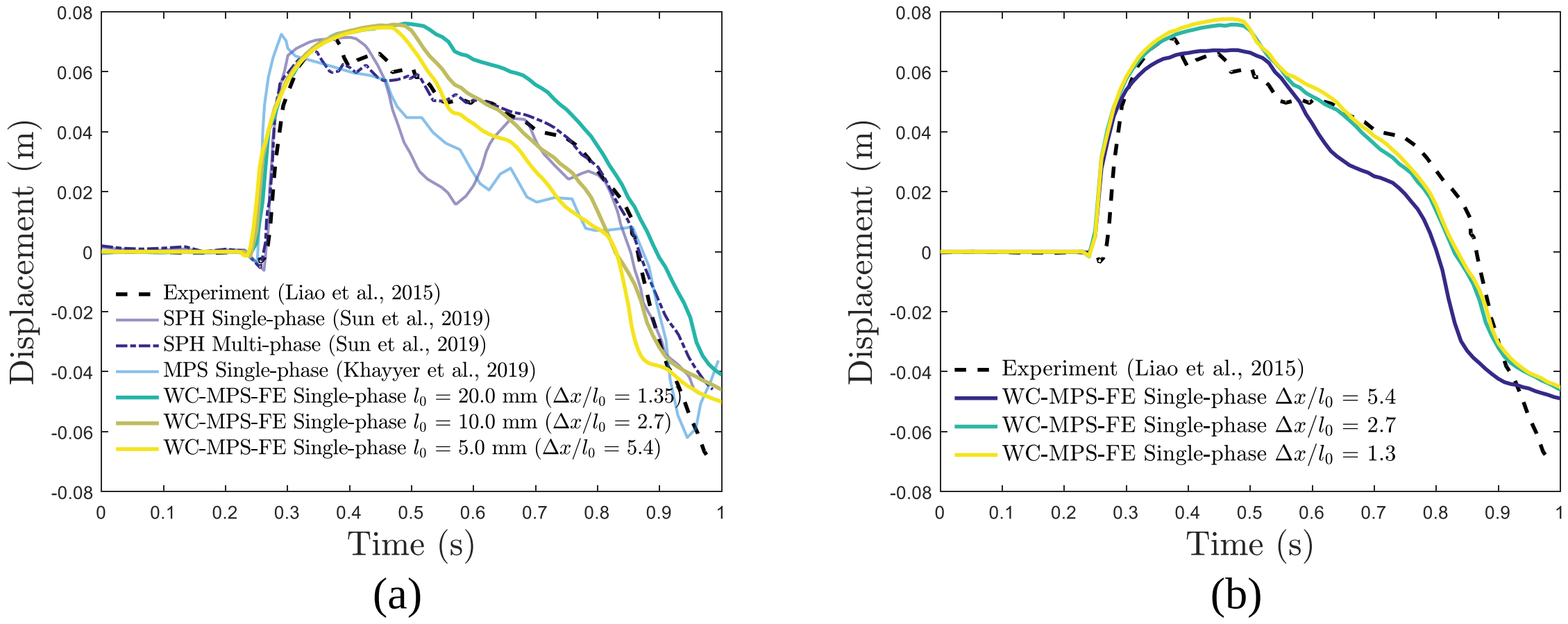

Fig. 3-12. Time series of the horizontal displacement of the free end of the elastic plate in the dam-break flow with an initial water column of height $H_F = 0.4$m. (a) Experimental data from Liao et al. [105], numerical results of single-phase and multi-phase SPH [23], single-phase MPS [22], and present single-phase WC-MPS-FE with $l_0 = 20, 10, 5$mm and $\Delta x = 27$mm. (b) Experimental data and present WC-MPS-FE with $l_0 = 10$mm and $\Delta x = 54, 27$ and 13.5mm ($\Delta x/l_0 = 5.4, 2.7$ and 1.3).

## 4 Concluding remarks

A 3D fluid-structure coupling between an enhanced weakly-compressible version of the MPS (CD-WC-MPS) and finite element (FE) methods, able to handle particles and finite elements of distinct sizes, was developed in the present work. The CD-WC-MPS was used to model violent free-surface flows while nonlinear thin-walled structures subject to large deformation were modeled by a geometrically exact shell approach.

Regarding the flow computation with CD-WC-MPS, the tuning-free diffusive term, introduced in the context of the MPS formulations, provides smooth and accurate pressure computations.

Discrete divergence operators were derived and applied for the explicitly represented polygon (ERP) model, which was also enhanced by introducing a repulsive Lennard-Jones force and a simple technique to avoid incorrect interactions between particles placed at opposite sides of thin shell solid modeling. As a result, the numerical stability of the fluid-solid coupling was improved.

Validation cases covering quasi-static and dynamic phenomena were carried out and the enhanced fluid solver reproduced well the analytical solutions and experimental measurements reported in the literature. Transient fluid-structure interaction (FSI) problems with violent free-surface and structures that undergo small or large rotations and displacements were simulated and the results demonstrated the robustness and reliability of the proposed coupled model. Stable and relatively accurate results were obtained even when a coarse particle resolution is considered. Furthermore, the CD-WC-MPS coupled with geometrically nonlinear shell allows for the simulations considering the relation between solid mesh size and particle distance larger than one, which is more cost-effective for modeling of large-scale problems of practical applications.

## 5 Acknowledgments

This study was financed in part by the Coordenação de Aperfeiçoamento de Pessoal de Nível Superior - Brasil (CAPES) - Finance Code 001. The second author acknowledges Conselho Nacional de Desenvolvimento Científico e Tecnológico (CNPq) under the grant 304680/2018-4.

## 6 Appendix A

To illustrate the accuracy of the present coupling scheme, we consider a hydrostatic water column of height $H_F = 0.2$m in a tank of height $H_T = 0.22$m and square elastic bottom of side length $L_p = 0.2$m, thickness $e_p = 0.002$m and clamped at all edges, according to Fig. 6-1. Simulations were performed for 2 seconds and three distance between particles, namely $l_0 = 20$, 10 and 5mm ($\Delta x/l_0 = 0.9$, 1.8 and 3.6), were evaluated. The physical properties and numerical parameters are summarized in Table 6.1 and Table 6.2, respectively. To achieve the static equilibrium, we adopted the dimensionless number $A_F = 1$, see Eq. (14), and a high structural numerical damping $\beta_R = 0.025$. The computed results are compared to the analytical solution of the central deflection $v_{MAX}$ given by [108]:

$$v_{MAX} = \alpha \frac{\rho_f g H_F L_P}{D_s}\ , \quad D_s = \frac{E_s e^3}{12(1-\nu_s^2)}\ , \quad \alpha = 0.00126\ . \tag{72}$$

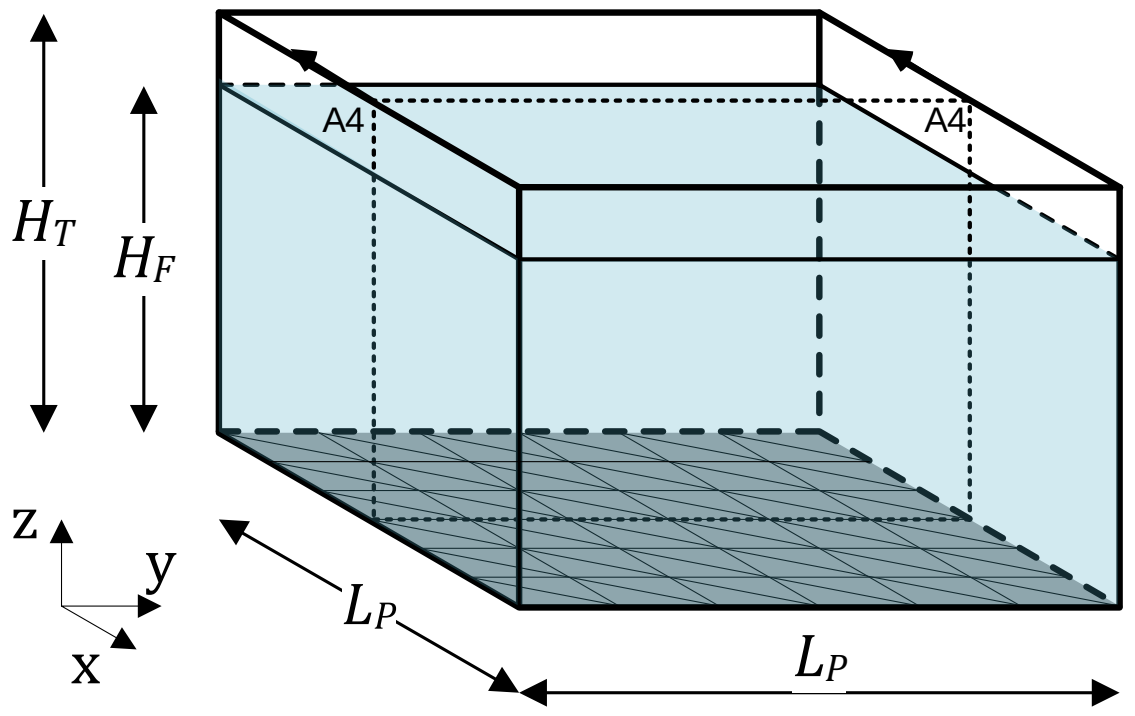

Fig. 6-1. Square plate of side length $L_p = 0.2$m, thickness $e = 0.002$m and clamped at all edges, under hydrostatic water column of height $H_F = 0.2$m. The section A4-A4 is the middle cross-sectional view.

Table 6.1. Square plate clamped at all edges and under hydrostatic water column. Physical properties of the fluid and elastic plate.

| Fluid | | Elastic plate | |
|---|---|---|---|
| Property | Value | Property | Value |
| Density $\rho_f$ (kg/m³) | 1000 | Density $\rho_S$ (kg/m³) | 1000 |
| Kinematic viscosity $\nu_f$ (m$^2$/s) | $10^{-6}$ | Young Modulus $E_S$ (Gpa) | 200 |
| | | Poisson ratio $\nu_s$ | 0.3 |
| | | Thickness $e_p$ (m) | 0.002 |

Table 6.2. Square plate clamped at all edges and under hydrostatic water column. Simulation parameters of the fluid and elastic plate.

| Fluid | | Elastic plate | |
|---|---|---|---|
| Parameter | Value | Parameter | Value |
| Particle distance $l_0$ (m) | 0.02, 0.01, 0.005 | Average element size $\Delta x$ (m) | 0.018 |
| Time step $\Delta t_F$ (s) | 5, 1.25, 1.25 $\times 10^{-4}$ | Time step $\Delta t_S$ (s) | $\Delta t_F$ |
| Effective radius $r_e$ (m) | $2.1\times l_0$ | Mesh elements | 16×16 |
| Sound speed $c_0$ (m/s) | 15 | Rayleigh damping stiff. Coef. | 0.025 |
| Courant number $C_r$ | 0.2 | Newmark coefficients $(\beta_N, \gamma_N)$ [90] | (0.3, 0.5) |
| Mach number $Ma$ | 0.1 | | |
| Dimensionless number $A_F$ | 1 | | |
| Dimensionless constant $C_{rep}$ | 1 | | |
| Surface thresholds $(\beta_{FS}, \varrho_{FS})$ | (0.98, 0.2) | | |

Fig. 6-2(a) illustrates the pressure field in the fluid adopting the initial particle distance $l_0 =$ 5mm and the displacement in the elastic plate, both behind the middle cross-sectional view A4-A4 (see Fig. 6-1), at the instant $t = 2.0$s. Given the intrinsic difficulties to obtain a static state with particle-based methods, the theoretical hydrostatic tank is numerically well-predicted.

Fig. 6-2(b) shows the evolution in time of the vertical displacements of the solid plate's central point. The plate presents a large oscillation until $t = 0.15$s in response to the suddenly change of the hydrostatic pressure field on the fluid particles. Approximately after $t = 0.2$s, the plate response becomes more stable and the computed displacement agrees very well with the analytical one, which is a good indication that the proposed coupling seems accurate. The computed results using the initial particle distances $l_0 = 20, 10$ and 5mm oscillate around the maximum displacements 0.0259 mm, 0.0265 mm and 0.0265 mm, respectively. Compared to the theoretical maximum displacement 0.027mm, the simulations using $l_0 \leq 10$mm show a small error of 1.85%, illustrating that the accuracy increases with the decrease of the particle distance, i.e., demonstrating the numerical convergence of the proposed model. The error was computed by:

$$\text{Error} = \left|\frac{\bar{X}_n - X_a}{X_a}\right|_{t\geq 0.5}, \tag{73}$$

where $X_a$ means the analytical result at the static equilibrium and $\bar{X}_n$ is defined here as the mean of the numerical result after reach the quasi-static equilibrium, namely $t \geq 0.5$s. The quantitative comparison between the analytical and present results is provided in Table 6.3.

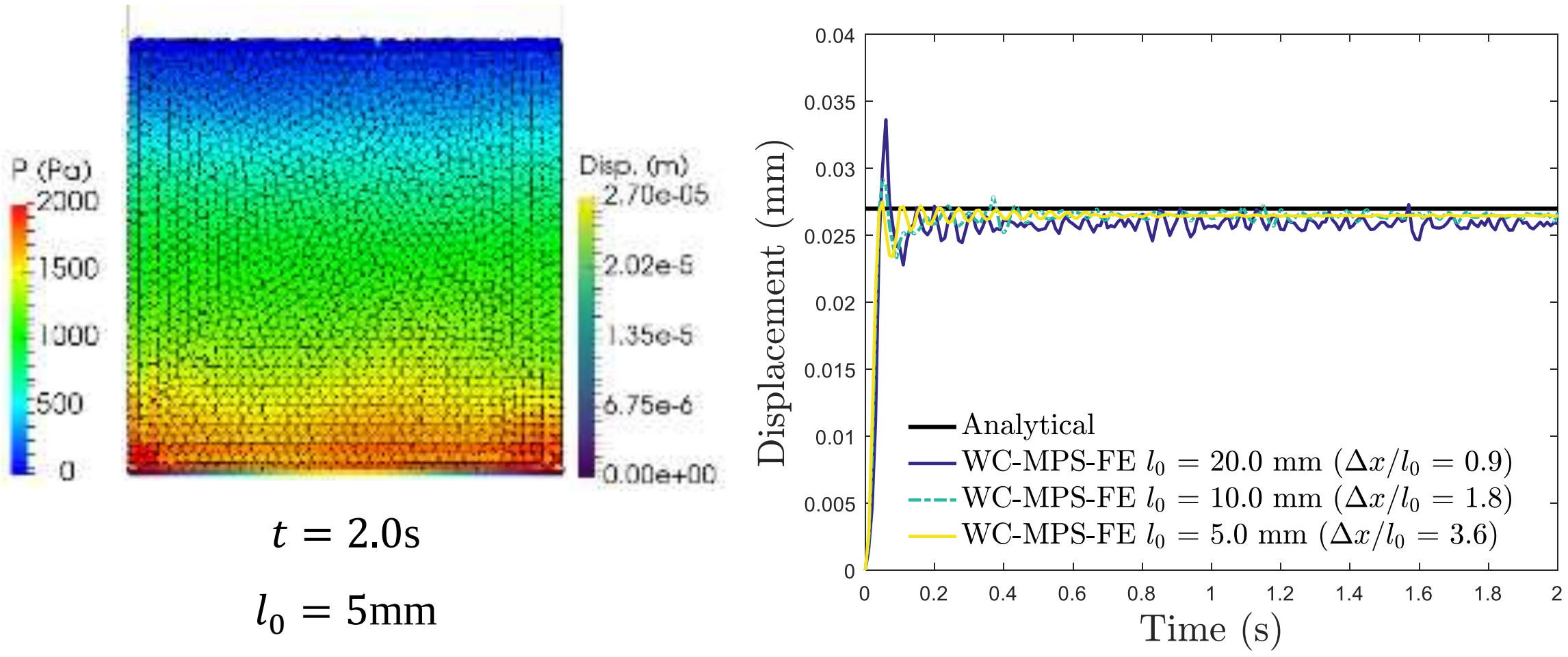

(a) (b)

Fig. 6-2. (a) Pressure field in the fluid using the particle distance $l_0 = 5$mm and the field of the displacement in the elastic plate at $t = 2.0$s (middle cross-sectional view A4-A4, see Fig. 6-1). (b) Time history of the plate's mid-span vertical displacement. Analytical and numerical results computed with the present WC-MPS-FE for different particle distances $l_0 = 20, 10, 5$mm and $\Delta x = 18$mm ($\Delta x/l_0 = 0.9, 1.8$ and $3.6$).

Table 6.3. Square plate clamped at all edges and under hydrostatic water column. Analytical and numerical results computed with the present WC-MPS-FE for different particle distances $l_0 = 20, 10, 5$mm.

| | Vertical displacement $v_{MAX}$ (mm) | Error (%) |
|---|---|---|
| Analytic solution | 0.0270 | - |
| WC-MPS-FE $l_0 = 20$mm ($\Delta x/l_0 = 0.9$) | 0.0259 | 4.07 |
| WC-MPS-FE $l_0 = 10$mm ($\Delta x/l_0 = 1.8$) | 0.0265 | 1.85 |
| WC-MPS-FE $l_0 = 5$mm ($\Delta x/l_0 = 3.6$) | 0.0265 | 1.85 |